\documentclass[a4paper,11pt]{article}
\pdfoutput=1 
\usepackage{float}
\usepackage{jcappub} 
\usepackage[T1]{fontenc} 
\title{\boldmath Weakly-Supervised Anomaly Detection in the Milky Way}

\author[a]{M. Pettee,}
\author[b]{S. Thanvantri,}
\author[a]{B. Nachman,}
\author[c]{D. Shih,}
\author[c]{\hspace{2cm} M.~R. Buckley,}
\author[d]{and J. H. Collins}

\affiliation[a]{\emph{Lawrence Berkeley National Laboratory, Berkeley, CA, USA}}
\affiliation[b]{\emph{University of California, Berkeley, Berkeley, CA,} USA}
\affiliation[c]{\emph{Rutgers University, New Brunswick, NJ, USA}}
\affiliation[d]{\emph{SLAC National Accelerator Laboratory, Menlo Park, CA, USA}}
\affiliation[e]{\emph{Bosch Research North America, Sunnyvale, CA, USA}}
\emailAdd{mpettee@lbl.gov}
\emailAdd{sowmya.thanvantri@berkeley.edu}
\emailAdd{bpnachman@lbl.gov}
\emailAdd{dshih@physics.rutgers.edu}
\emailAdd{mbuckley@physics.rutgers.edu}
\emailAdd{jcollins@slac.stanford.edu}

\abstract{Large-scale astrophysics datasets present an opportunity for new machine learning techniques to identify regions of interest that might otherwise be overlooked by traditional searches. To this end, we use Classification Without Labels (\textsc{CWoLa}), a weakly-supervised anomaly detection method, to identify cold stellar streams within the more than one billion Milky Way stars observed by the {\it Gaia} satellite. \textsc{CWoLa} operates without the use of labeled streams or knowledge of astrophysical principles. Instead, we train a classifier to distinguish between mixed samples for which the proportions of signal and background samples are unknown. This computationally lightweight strategy is able to detect both simulated streams and the known stream GD-1 in data. Originally designed for high-energy collider physics, this technique may have broad applicability within astrophysics as well as other domains interested in identifying localized anomalies.}

\begin{document}
\maketitle
\flushbottom

\section{Introduction}
\label{sec:intro}

\subsection{Motivation}

The history of our home galaxy, the Milky Way, has been marked by the ongoing aggregation of stars, gas, and dark matter from various sources throughout the Universe. These accumulation events include mergers with other galaxies as well as smaller-scale gravitationally-bound groupings of stars such as globular clusters. Though many such collisions occurred in the distant past, lingering remnants from more recent collisions contain crucial information revealing the Milky Way's merger history \cite{Johnston_1998, helmi, Helmi_2018, Belokurov_2006, Belokurov_2018, Malhan_2021}, underlying gravitational potential \cite{Law_2010, eyre, Dehnen_2004, kamdar2021stellar, reino, Nibauer_2022}, and dark matter content \cite{Carlberg_2012, Sanders_2016, erkal, Bonaca_2019, Banik_2019, Bonaca_2020, Purcell_2012, Necib_2019}.

Since 1971, astronomers have observed collections of stars called \emph{stellar streams}: thin, ribbon-like arcs orbiting the Milky Way's galactic center \cite{Eggen_1971}. These dynamically cold associations of stars are thought to be the result of gravitational tidal forces from the Milky Way disrupting and warping nearby low-mass progenitors -- dwarf galaxies or globular clusters -- until the stars are no longer gravitationally self-bound. Due to their shared origin, the stars tend to share many characteristics ranging from proper velocity to age. 

To date, nearly 100 stellar streams have been identified in the Milky Way \cite{Mateu_2023}. They are challenging to discover and study due to their sparse densities and wide angular extents. Gaining a more detailed understanding of the structure of known streams as well as uncovering additional streams will be critical for deepening our understanding of the Milky Way. An extensive catalogue of high-precision stream measurements would greatly improve our estimation of the particularites of Galactic large-scale and small-scale structures, including contributions from cold dark matter.  

\subsection{Related Work}
Traditional methods of identifying stellar streams look for groupings of stars that are similar along various metrics: color and magnitude \cite{Rockosi_2002, matchedfilter}, velocity \cite{Williams_2011, Arifyanto_2006, Duffau_2005}, or position along great circle paths across the sky \cite{Johnston_1996, Mateu_2017}. More recently, an automated technique called \textsc{Streamfinder} \cite{streamfinder}, leverages both position and kinematic information to construct volumes called ``hypertubes'' in a multidimensional phase space around a stream candidate, then iterates to optimize a statistic similar to a log-likelihood to determine the best parameters for a given hypertube. While this algorithm has led to the discovery of multiple low-density stellar streams, it makes several assumptions based on astrophysical principles. For instance, it assigns the distance to a stream based on a particular choice of isochrone and assumes a specific Milky Way potential to calculate a stream's orbit. Data mining and clustering techniques such as DBSCAN have also been applied to \textit{Gaia} data for stellar stream searches \cite{Borsato_2019}.  

Machine learning techniques have also been deployed in search of stellar streams. Recently, an unsupervised machine learning technique called \textsc{Via Machinae} \cite{viamachinae, viamachinae2}, based on the \textsc{ANODE} method~\cite{Nachman:2020lpy} originally designed for high-energy particle physics searches, was applied towards the automated discovery of stellar streams within the {\it Gaia} Data Release 2 catalogue. \textsc{Via Machinae} combines a normalizing flow density estimation technique~\cite{https://doi.org/10.48550/arxiv.1505.05770} for anomaly detection with a line-finding algorithm to identify stellar stream candidates without the use of detailed assumptions about isochrones or stream orbits.

\subsection{\textsc{CWoLa}: Classification Without Labels}

Machine learning has been widely and successfully applied in the context of fundamental physics to the classification and description of various physical phenomena ranging from subatomic to cosmological scales. These techniques excel at identifying complex patterns in a dataset without imposing any prior assumptions about their distributions. When dealing with real data with partially inaccurate or incomplete labels, weakly-supervised machine learning methods can be particularly helpful. 

The fields of high-energy particle physics and astrophysics share a common interest in identifying localized features, meaning overdensities of data concentrated in contained regions of phase space, within vast and high-dimensional datasets. Model-independent forms of anomaly detection -- the process of identifying these localized features that deviate from a dataset's typical characteristics -- can efficiently filter these large datasets in an unbiased manner and aid in potential discoveries.

In this paper, we demonstrate the first astrophysical application of \emph{Classification Without Labels} (\textsc{CWoLa}\footnote{Note: \textsc{CWoLa} is pronounced ``koala''.}) \cite{cwola}, a weakly-supervised machine learning technique based on a simple, lightweight neural network classifier. Like \textsc{ANODE} \cite{Nachman:2020lpy}, the neural network architecture used in \textsc{Via Machinae} \cite{viamachinae, viamachinae2}, \textsc{CWoLa} was originally designed for identifying particles within high-energy particle physics datasets. As such, \textsc{CWoLa} has been applied as a promising model-agnostic anomaly detection method for searching for localized features such as the potential signatures of new fundamental particles at the Large Hadron Collider (LHC) \cite{cwola_hunting,Collins:2018epr,ATLAS:2020iwa}, but until now has not been used on astrophysics datasets.

This analysis uses the same datasets as in \textsc{Via Machinae} and follows the same general framework. We scan across regions defined by one proper motion coordinate and train neural networks to assign an anomaly score to stars in these regions using five input variables: two angular position coordinates, one proper motion coordinate (the one not used for the scan), magnitude, and color. In both analyses, we scan along proper motion because we expect stellar streams to be kinematically cold and therefore relatively localized along this feature compared to background stars, though we do not make any requirements that the streams fall on a particular orbit or stellar isochrone. Once the anomaly scores are assigned, we look at the subsets of stars from each scanning window with the highest anomaly scores, cluster them, and apply some fiducial selections to further refine them. We apply these techniques on a known stream called GD-1 as well as simulated stellar streams as benchmarks. 

Despite these overarching similarities, this analysis also departs from that of \textsc{Via Machinae} in several ways. Most significantly, the choice of neural network model to assign anomaly scores and the post-training clustering method are different. These alternative choices in methodology have been implemented with the aim of achieving similar performance in anomaly detection with a much more computationally lightweight framework. A detailed discussion of similarities, differences, and potential benefits or drawbacks between these analyses is presented in Section \ref{sec:discussion}. 

\subsection{Outline}

This paper is organized as follows. First, in Section \ref{sec:gaia}, we describe the {\it Gaia} dataset and how it is processed for use in our anomaly detection studies. Then, in Section \ref{sec:cwola}, we explain the methodology of applying \textsc{CWoLa} for anomaly detection and our particular implementation of \textsc{CWoLa} on {\it Gaia} data, including how we define the signal and sideband regions as well as the neural network model architecture and training procedure. The results of applying \textsc{CWoLa} to the known stellar stream GD-1 are listed in Section \ref{sec:results}. Finally, in Section \ref{sec:discussion} we conclude with a discussion of \textsc{CWoLa}'s potential usefulness in aiding future stellar stream discoveries and some further steps for this work that can bring us closer to that goal. 

\section{{\it Gaia} Dataset}
\label{sec:gaia}

\begin{figure*}[ht!]
    \centering
    \includegraphics[width=\textwidth]{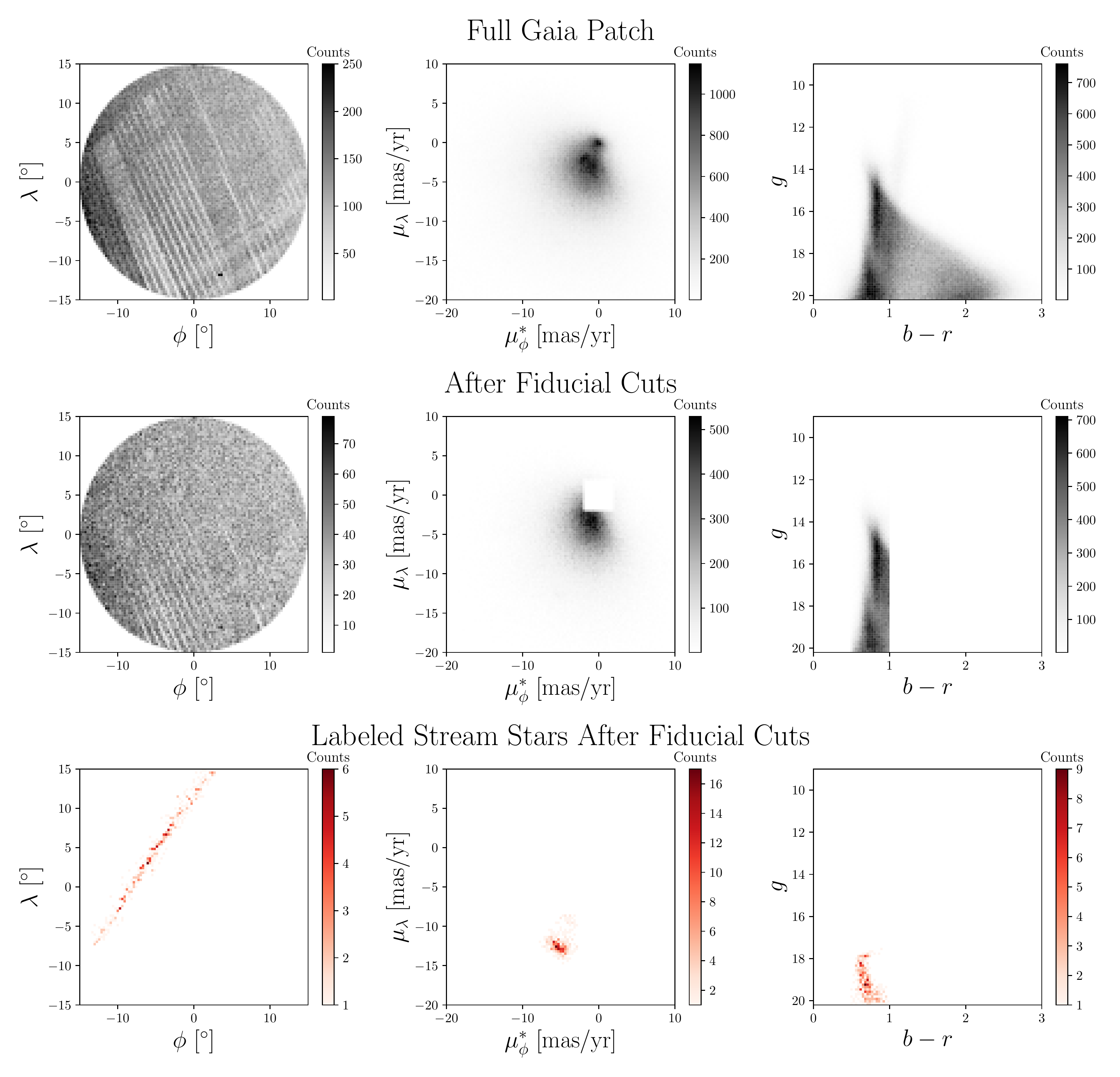}
    \caption{Two-dimensional histograms of the six features used in this analysis are illustrated for a single patch in the sky containing some GD-1 stars. This patch is centered at ($l=207.0$, $b=50.2$). The top row shows the full patch with no selections applied. The second row shows the patch with fiducial selections applied: $g < 20.2$ to reduce streaking; $|\mu_{\lambda}| > 2$  mas/year or $|\mu_\phi^{*}| > 2$ mas/year to remove too-distant stars, and $0.5 \leq b-r \leq 1$ to focus on identifying cold stellar streams. The third row indicates the six features for the GD-1 stream following the fiducial selections.}
    \label{fig:inputs}
\end{figure*}

The {\it Gaia} catalogues are extensive astronomical data releases mapping the stars populating the Milky Way \cite{gaia_mission}. The {\it Gaia} satellite itself was launched in 2013, and its data catalog is being released in discrete stages throughout its operational lifespan through 2025. The data analyzed for this work comes from {\it Gaia}'s Data Release 2 (DR2), the collection of {\it Gaia}'s observations from July 25, 2014 to May 23, 2016 \cite{gaia_dr2}. {\it Gaia} DR2 contains position, proper motion, and photometric information for approximately 1.3 billion stars, representing around 1\% of the total star population of the Milky Way. While this analysis was already in process, {\it Gaia} released two additional data releases: Early Data Release 3 (eDR3) and Data Release 3 (DR3). Were this analysis to be extended using eDR3 or DR3 data, we would likely see some further improvements due to reduced measurement errors. Other changes in eDR3 and DR3 include improved distance and radial velocity measurements for a small subset of stars, but these variables are not considered in this analysis.

While {\it Gaia} DR2 also contains information on parallax and mean radial velocity, these variables are not considered as input variables to \textsc{CWoLa} for this analysis. We exclude parallax because it is not as reliable a feature as our other observables for stars as distant as the stream members in which we are most interested. However, we do use parallax to apply a cut on the \emph{Gaia} data to restrict stars to a maximum parallax of 1, meaning stars at a distance of at least 1 kpc. Mean radial velocity, or motion along the axis between the Earth and each star, is also measured for some stars in the {\it Gaia} dataset, but {\it Gaia} DR2 contains radial velocities for only about 7 million stars, representing less than a percent of the overall star catalog. We therefore omit radial velocity in order to maximize available training statistics. 

As in \cite{viamachinae}, we use the stellar stream GD-1, discovered in 2006 \cite{gd1}, as the main candidate for evaluating the performance of \textsc{CWoLa} as a stream-finding technique. Most likely the remains of a tidally-disrupted globular cluster, GD-1 consists of primarily metal-poor stars totaling approximately $2\times 10^4\ M_\odot$ \cite{gd1_info}. It lies at a distance of approximately 10 kpc from the Sun and 15 kpc from the Galactic Center. GD-1 is especially narrow \cite{gd1}, dynamically cold \cite{deboer}, and bright \cite{erkal} compared to other stellar streams in the Milky Way. It also contains various known physical peculiarities including gaps and wiggles \cite{gap}, offshoots (``spurs'') and overdensities (``blobs'') \cite{Bonaca_2019}, and even a surrounding ``cocoon'' of stars \cite{Malhan_2019}. The faithful reconstruction of some of these density perturbations can therefore serve as an additional metric indicating the physical validity of \textsc{CWoLa}'s outputs.

\subsection{Data Preprocessing}
We use the same dataset of {\it Gaia} stars as in Ref. \cite{viamachinae}. Following the same data processing methodology, we train our model on a series of 21 overlapping circular ``patches'' of the {\it Gaia} dataset with radius 15$^{\circ}$. While the natural angular position coordinates in DR2 are right ascension ($\alpha$) and declination ($\delta$), we use rotated and centered coordinates $\phi$ and $\lambda$ (as well as rotated proper motion coordinates $\mu_\phi$ and $\mu_\lambda$) such that each patch has a Euclidean distance metric and is centered at $(\alpha_0, \delta_0) = (0\text{\textdegree},0\text{\textdegree})$. This transformation is performed using \textsc{Astropy} \cite{astropy:2013,astropy:2018,astropy:2022}. Each patch's center location is also documented in \cite{viamachinae}.

Beyond these two rotated and centered angular positions, we consider four additional features associated with each star in the dataset: two angular proper motions ($\mu_{\phi *}$, where $\mu_{\phi *}\equiv\mu_\phi\text{cos}(\lambda)$, and $\mu_\lambda$), color ($b-r$, where $b$ represents the brightness of the blue photometer and $r$ represents the brightness of the red photometer), and magnitude ($g$). Distributions of the six relevant variables used in the analysis ($\phi$, $\lambda$, $\mu_{\phi^*}$, $\mu_\lambda$, $b - r$, and $g$) are shown for one such patch in Figure~\ref{fig:inputs}. These patches cover an irregularly-shaped region stretching between approximately $\alpha \in [-250^{\circ}, -100^{\circ}]$ and $\delta \in [-10^{\circ}, 80^{\circ}]$. 

Following these selections, 1,957 of the approximately 8 million total stars considered for this analysis are labeled as likely belonging to the GD-1 stream using the catalogues developed by Refs. \cite{Price_Whelan_2018, pwb18}. This choice of labeling is based on selections in position, proper motion, and along an isochrone in color and magnitude. While these labels cannot be considered fully accurate or complete, they serve as a helpful reference for evaluating our model's efficacy. 

\section{Methods}
\label{sec:cwola}
\subsection{Classification Without Labels (\textsc{CWoLa})}
\textsc{CWoLa} is a weakly-supervised machine learning technique designed to find anomalous features in a dataset that are localized along at least one dimension. It was originally designed for applications within high-energy particle physics, where mixures of particle classes with unknown proportions of signal and background are common. It detects anomalies by scanning along a localized dimension -- for instance, the invariant mass of the final state of a particle physics even -- and learning to distinguish between mixtures of data classes where the precise class proportions within each mixture need not be known. Simply by learning to differentiate regions with higher vs. lower proportions of signal, i.e. ``signal'' vs. ``sideband'' regions, \textsc{CWoLa} can be a powerful indicator of patterns of anomalous events. 

\begin{figure*}[t!]
    \centering
    \begin{subfigure}[t]{0.55\textwidth}
        \centering
        \includegraphics[width=\textwidth]{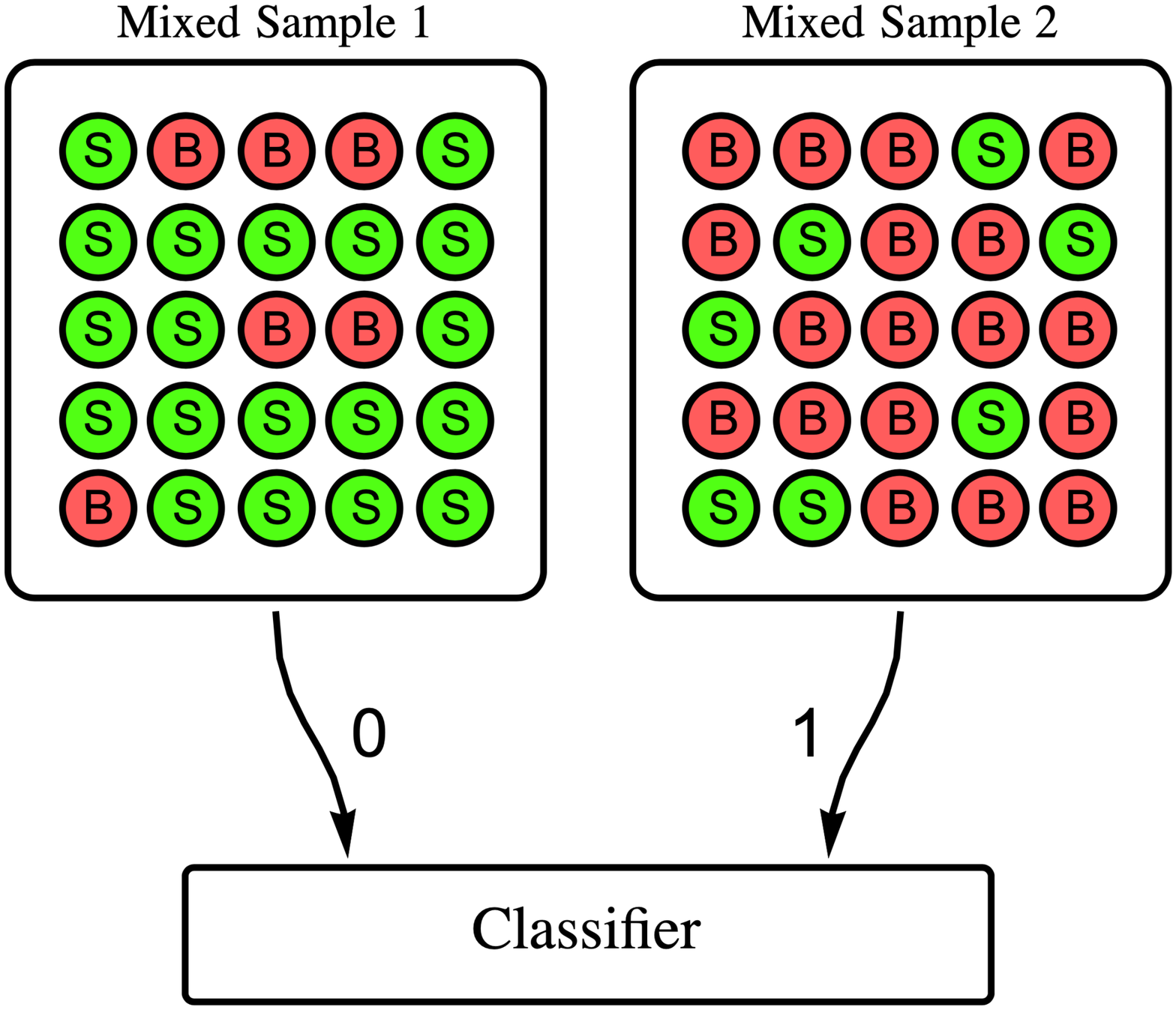}
        \caption{Signal-enriched ($M_1$) vs. Signal-depleted ($M_2$)}
        \label{fig:sig_enriched}
    \end{subfigure}%
    \begin{subfigure}[t]{0.45\textwidth}
        \centering
        \includegraphics[width=0.95\textwidth]{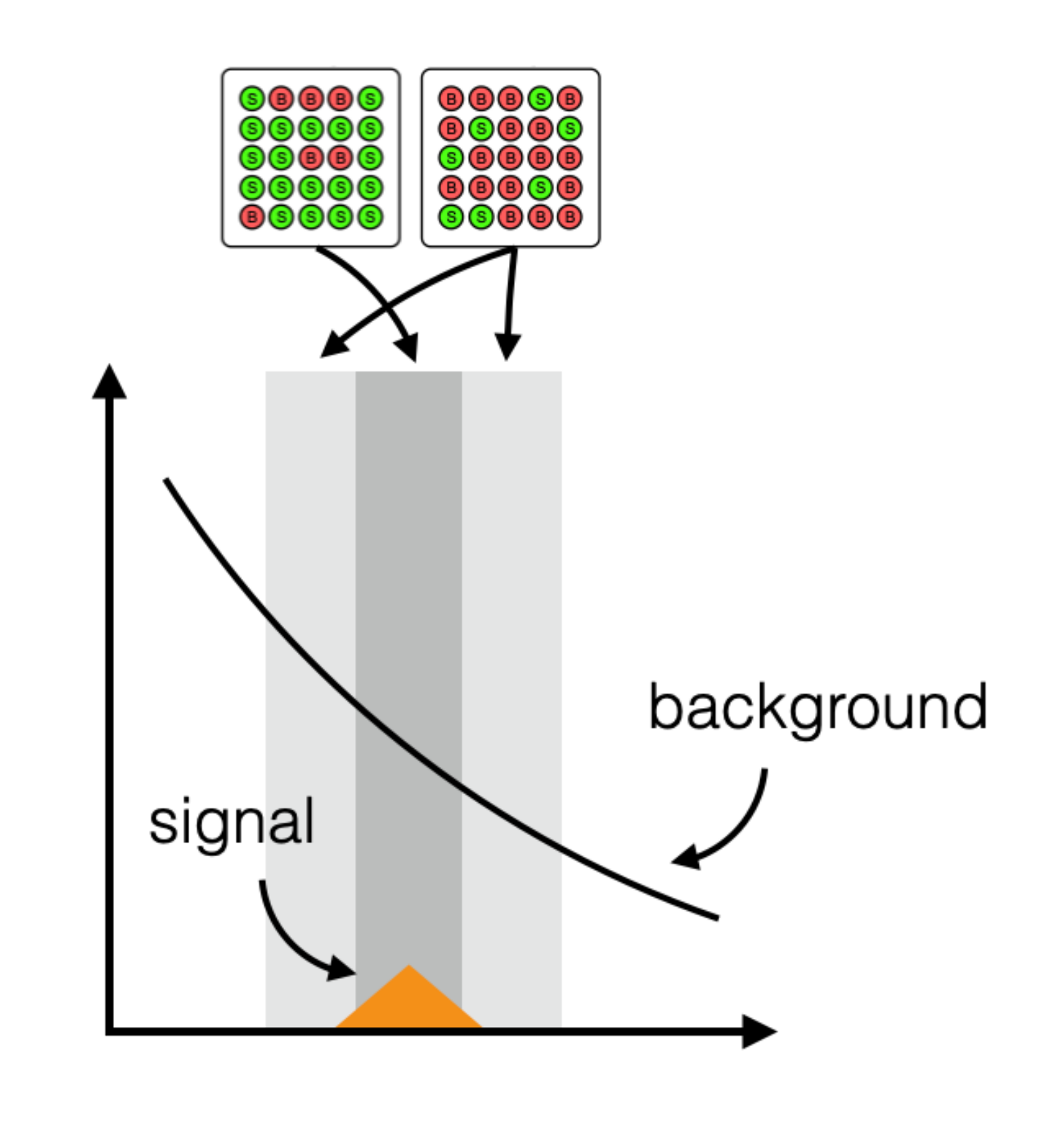}
        \caption{A localized anomaly}
        \label{fig:localized_anomaly}
    \end{subfigure}
    \caption{Signal-enriched and signal-depleted groups are pictured above. The green data points labeled ``S'' represent signal events, while the red data points labeled ``B'' represent background events. The signal and sideband regions are chosen such that more signal events (shown in orange) are located in the central signal region than the surrounding sideband region.}
\end{figure*}

Consider a signal-enriched mixture $M_1$ and a signal-depleted mixture $M_2$, as shown in Figure~\ref{fig:sig_enriched}. ``Signal'' refers to an object class of interest -- here, a member of a localized anomalous feature, such as a stellar stream, that one would like to detect -- while ``background'' refers to objects not part of the anomaly. Both mixtures contain signal and background events, but the signal-enriched mixture has significantly more signal events relative to the signal-depleted mixture (i.e., $f_1 > f_2$, where $f_i$ indicates the fraction of signal events in each mixture). We exploit the fact (see proof of Theorem 1 in \cite{cwola}) that an optimal classifier trained to distinguish events between $M_1$ and $M_2$ is the same as an optimal classifier trained in a fully-supervised manner to distinguish signal from background events. Importantly, the exact proportions of signal in each mixture ($f_1$ and $f_2$) need not be known for this to hold. This theorem relies on the Neyman-Pearson lemma \cite{neyman} that states that an optimal classifier $h(\vec{x})$ is any function monotonic to the likelihood ratio constructed from the probability distributions of signal and background $p_S(\vec{x})$ and $p_B(\vec{x})$ for input variables $\vec{x}$.

We can apply \textsc{CWoLa} as a model-agnostic, data-driven anomaly detection technique \cite{cwola_hunting,Collins:2018epr} by identifying a certain feature of our dataset that might contain a localized anomaly, as illustrated in Figure~\ref{fig:localized_anomaly}. We then train a fully-supervised classifier to distinguish between events from a ``signal region'' and a surrounding ``sideband region'', as defined by ranges of this feature. The inputs to this classifier are auxiliary variables that should be decorrelated from the characteristic used to define the signal and sideband regions if no anomaly is present. If an anomaly is present and contained primarily in the signal region, then we expect the anomalous events to be ranked more highly by the classifier. We can then repeat this process by sliding the signal and sideband windows across a range of values. For each choice of signal and sideband, we apply a threshold on the classifier output score (i.e. the top $N$ events or a top percentile of the test set) such that only the highest-score events remain. When an anomaly is present, these highest-score events will contain an enhanced signal-to-noise ratio of events.

\begin{figure}[t]
\centering
\includegraphics[width=\textwidth]{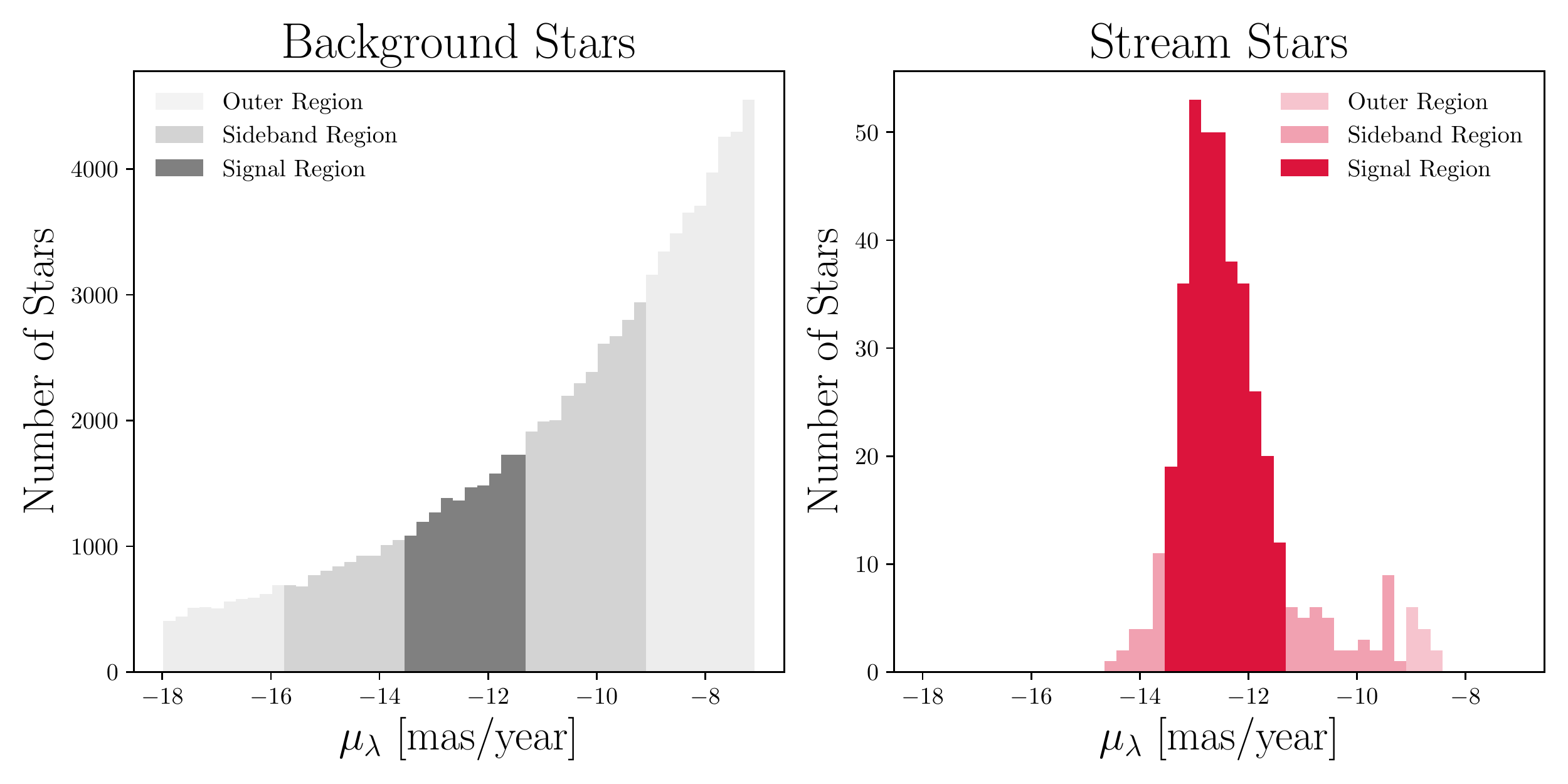}
\caption{\label{fig:mu_delta} Stars associated with the stellar stream GD-1 are highly localized in $\mu_\lambda$ space in comparison with background stars for the same patch of {\it Gaia} data seen in Figure \ref{fig:inputs}. The signal region, shown in the darkest regions in each plot, is defined by taking $\pm1\sigma$ from the median $\mu_\lambda$ value for the stream stars, which in this case is $[-13.6, -11.4]$. The sideband region is defined by taking $\pm3\sigma$ from the stream's median $\mu_\lambda$ value, excluding the signal region: $[-15.8, -13.6)$\ \&\ $(-11.4, -9.3].$}
\label{fig:mu_lambda}
\end{figure}

\begin{figure}
    \centering
    \includegraphics[width=\textwidth]{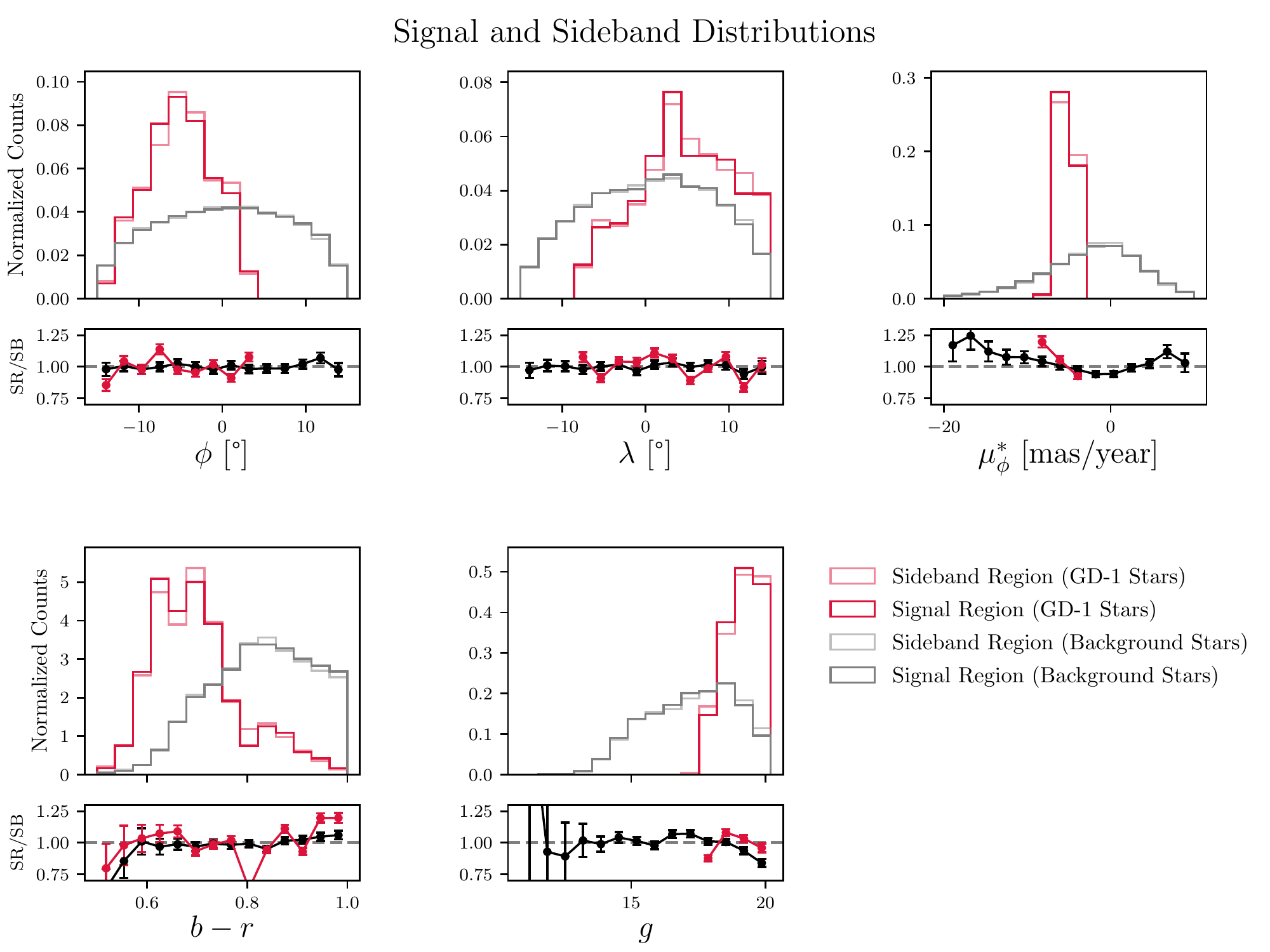}
    \caption{Distributions for the five neural network inputs are compared for both GD-1 stars (in red) and background stars (in grey) across signal and sideband regions. The patch shown here is the same example patch from Figure \ref{fig:inputs}. For both stream and background stars, the distributions for these five variables across the signal and sideband regions are approximately indistinguishable.}
    \label{fig:sr_sb_equiv}
\end{figure}

The \textsc{CWoLa} anomaly search makes two key assumptions during its procedure. First, it requires that the anomaly is localized in the dimension over which we search. Because stellar streams are kinematically cold, they are relatively localized in both coordinates of proper motion. We select $\mu_\lambda$ as the primary coordinate used to define the signal and sideband regions for each patch of DR2. A histogram demonstrating the highly localized nature of $\mu_\lambda$ within an example patch of GD-1 is shown in Figure~\ref{fig:mu_delta}. Second, it expects that background and signal events are indistinguishable between the signal and sideband regions. This is a reasonable assumption, as shown for an example patch of data in Figure \ref{fig:sr_sb_equiv}. 

\subsection{Defining Signal \& Sideband Regions}

For each of the 21 patches of DR2 considered in this study, we construct a signal region to contain the bulk of the stream stars available and neighboring sideband regions such that the background stars in both signal and sideband regions are as close to indistinguishable as possible. Ideally, the stars in both regions should have similar characteristics: background stars in the signal region should closely resemble background stars in the sideband region, and same logic applies for the stream stars. Stars belonging to a stellar stream make up a small fraction of total stars within each patch, so in our case, each signal region will still be dominated by background stars not labeled as belonging to the stream. However, each signal region should have a higher signal-to-background ratio than the sideband region. 

Signal regions are ideally defined by a range of $\mu_\lambda$ values that encompass the bulk of the stream stars. There are many valid ways to define these regions in general, and in some cases, the best definitions may be model-dependent. 

For this proof-of-concept result, we opted for idealized signal and sideband limits based on where we know the stream to be concentrated in proper motion. However, it is crucial to note that for a full-scale anomaly search, one could scan across a range of $\mu_\lambda$ values, meaning one should still be sensitive to streams even if different signal and sideband region were selected. 

In this case, we define the signal region in each patch as the region within one standard deviation of the median $\mu_\lambda$ of the GD-1 stars in the patch. The sideband region within each patch is then defined as the stars falling within $[-3\sigma, -\sigma]$ or $[\sigma, 3\sigma]$ of the median. Given that the signal region encompasses a bulk of the stream, the sideband regions will have significantly fewer stream stars and will be signal-depleted, as desired. In practice, the average span of the signal regions across the 21 patches in $\mu_\lambda$ was $2.34\pm 0.36$ mas/year and the average span of the sideband regions was $7.02\pm 1.09$ mas/year.

\subsection{Neural Network Architecture and Training Procedure}
We implement \textsc{CWoLa} with a neural network built in Keras \cite{keras} with a TensorFlow backend  \cite{tensorflow}. The model consists of 3 hidden fully-connected layers, each with a layer size of 256 nodes and a ReLU activation \cite{relu}. Each fully-connected layer is followed by a dropout operation with a dropout rate of 20\% \cite{dropout}. These layers are followed by a final output layer of a single node with a sigmoid activation. Hyperparameter values for layer size, batch size, and number of $k$-folds (described below) was chosen via an optimization using Optuna \cite{optuna}.

For each of the 21 {\it Gaia} patches considered in our search for GD-1, we train a series of classifiers to separate stars labeled as part of the signal region from stars labeled as part of the sideband region. This quality defines \textsc{CWoLa} as being ``weakly-supervised'': it operates with a little more information than a fully unsupervised network, as we expect an optimal signal region to contain a higher fraction of GD-1 stars than in the sideband region, but it does not have access to the actual GD-1 labels. The training procedure, which closely aligns with other \textsc{CWoLa} searches \cite{cwola_hunting, Collins:2018epr}, unfolds as follows:

\begin{enumerate}
    \item \textbf{$k$-folding:} We implement stratified $k$-folding ($k$ = 5) to randomly divide all the stars in a given patch into five sections, or ``folds''. Each fold is chosen such that the overall percentage of stars labeled as ``signal'' vs. ``sideband'' is also maintained within each fold. The first fold (20\% of all stars) is reserved as a test set. The second fold (another 20\% of all stars) is used as a validation set. The remaining 60\% of stars are used for training.
    \item \textbf{Train:} Next, a classifier with the architecture specified above is trained on the training set with a batch size of 10,000 for up to 100 epochs, though early stopping with a patience of 30 typically halts the training process well before this limit. The large batch size is necessary due to the low number of labeled stream stars in the overall dataset -- for example, one patch on the tail end of GD-1 has a stream star population of just 0.15\%. Large batch sizes therefore help ensure that more than a handful of stream stars will be seen at a time by the network during training. We use the binary cross-entropy loss function and Adam optimizer \cite{adam}. The validation set is used to monitor the validation loss for early stopping. 
    \item \textbf{Repeat:} The classifier training is repeated twice more, each time with a random initialization of trainable parameters. Of the three distinct trainings, the weights are stored for the model with the lowest validation loss. 
    \item \textbf{Cycle through validation sets:} This process is repeated using each of the remaining folds as a validation set with the exception of the test set, which remains unchanged. For each configuration, the remaining three folds besides those used for the test and validation sets are used for training. 
    \item \textbf{Evaluate on test set:} Each of the best models trained using the four $k$-fold options for the validation set is evaluated on the test set. The final \textsc{CWoLa} score for each star in the test set is defined as the average across the four scores. 
    \item \textbf{Combine test sets:} This entire process is repeated, cycling through each of the five possible $k$-folds as a test set. These test sets are then concatenated into a single dataset such that every star in the patch ends up in the test set exactly once. 
\end{enumerate} 

\vfill
\clearpage

\subsection{Model Evaluation}
After training the neural network classifiers inherent in the \textsc{CWoLa} methodology, a series of fiducial selections are applied to each patch to further refine the results and optimize for the highest possible signal-to-noise ratio. The fiducial selections used are almost identical to their counterparts in \cite{viamachinae, viamachinae2}: 
\begin{itemize}
    \item $g < 20.2$, to ensure uniform acceptance by the {\it Gaia} satellite
    \item $|\mu_{\lambda}| > 2$  mas/year or $|\mu_\phi^{*}| > 2$ mas/year, to remove very distant stars that are predominantly concentrated near zero proper motion and therefore not equally distributed throughout the patch
    \item $0.5 \leq b-r \leq 1$, to isolate old and low-metallicity stellars streams in color space.
\end{itemize}

Unlike in \textsc{Via Machinae}, however, we do not need to apply a cut restricting the patch radius from 15\textdegree $\rightarrow$ 10\textdegree\ after training, as \textsc{CWoLa} is not influenced by the edge effects seen in the \textsc{ANODE} implementation. 

\textsc{Via Machinae} employs a sophisticated line-finding strategy using modified Hough transforms \cite{hough} to search for line-like structures in the identified anomalous stars and then combines these line segments into an overall stream candidate. We achieve similar results with a relatively lighter computational load using $k$-means clustering \cite{kmeans} ($k=2$) in proper motion space. Following the grouping of stars into two clusters, we select the cluster with the largest population of stars and discard the stars in the other cluster. This is motivated by our expectation that the stellar stream should be kinematically cold, therefore the velocities of its constituent stars should be densely clustered in velocity space. This clustering strategy is likely best used as a post-discovery tool and may not perform well in contexts with high numbers of contaminant stars not belonging to a stream. In these cases, opting for a larger $k > 2$ or a line-finding technique could instead be a better choice.

Following these fiducial selections, model performance was evaluated by applying the classifier to stars in the combined test set equivalent to the entire patch. The output scores were sorted from highest to lowest, where higher values indicated that the model ranked those stars as more likely to belong to the signal region than the sideband region. The top $N=250$ stars, ranked by neural network output scores, are chosen for evaluation. 

The number 250 was chosen following an optimization for both purity (percentage of top-ranked stars overlapping with labeled GD-1 stars) and completeness (percentage of labeled GD-1 stars covered by \textsc{CWoLa}'s top-ranked stars). In principle, however, one could isolate a different absolute number or relative percentage of top stars, though it would be advisable to stay under the average of 430 labeled stream stars per patch.

It is worth emphasizing that this method of model evaluation requires ground truth labels. In the absence of reliable stream labels, or in the case of discovering a new stream, we must employ different methods to evaluate model performance, not to mention a modified strategy for the model implementation itself. We discuss this further in Section~\ref{sec:discussion}.

\clearpage

\section{Results}
\label{sec:results}

Before looking at real {\it Gaia} data, we also evaluated the performance of \textsc{CWoLa} when applied to 100 randomly-chosen simulated stellar streams. As shown in Appendix \ref{appendix:mock}, with just two passes of \textsc{CWoLa}, 96\% of streams are identified with nonzero purity, of which 69\% are identified with a purity larger than 50\%. 

\subsection{GD-1 Stream Identification}

\begin{figure}
\centering 
\includegraphics[width=\textwidth]{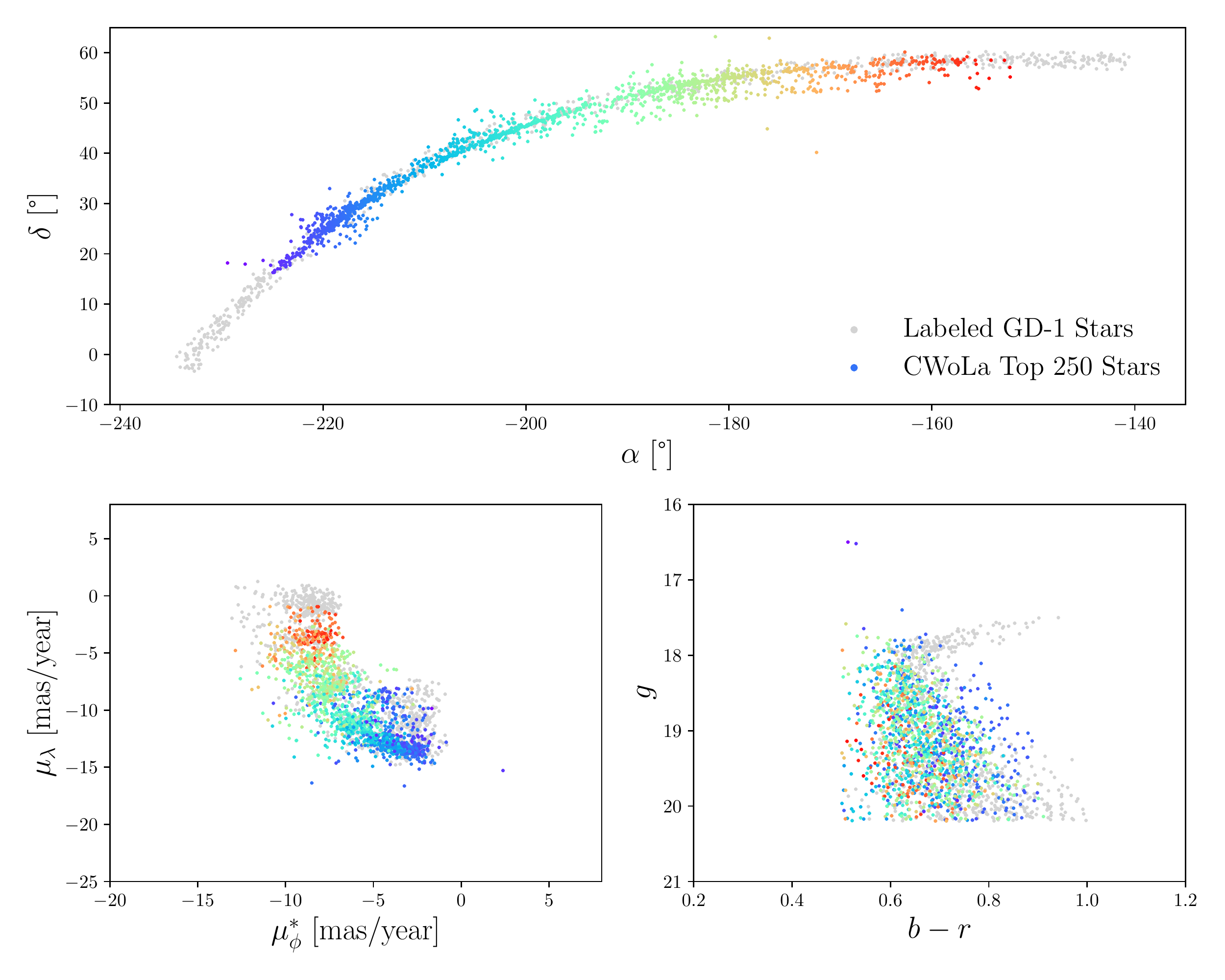}
\caption{\label{fig:rainbow} The full scope of stars identified by the \textsc{CWoLa} method in overlapping patches across the angular range corresponding to GD-1. Light gray dots indicate the ground truth labeling of GD-1 stars \cite{pwb18}, while the top 250 stars identfied by \textsc{CWoLa} in each patch are indicated in colored dots. The colors are chosen to correlate with each star's $\alpha$ value.}
\end{figure}

The combined results of applying the \textsc{CWoLa} technique to each of the 21 patches of {\it Gaia} DR2 are shown in Figure \ref{fig:rainbow}. Results for individual patches are detailed in Appendix \ref{appendix:patches}. Across the 21 patches, 1,498 unique GD-1 stars pass our fiducial selections. 1,360 unique stars are identified in the combined top $N=250$ stars for each CWoLa patch. Of these, 760 are part of the labeled GD-1 star set \cite{pwb18}. Thus, across the entire stream, we achieve a purity of 56\% and a completeness of 51\%. In our optimization studies, we found that stream purity plateaued at a maximum value of 78\% using the top $N=25$ stars in each patch, but this choice of $N$ only yields a competeness of 13\%. Conversely, choosing $N=300$ yields a reduced purity of just $30\%$, but a higher completeness of 54\%.
\enlargethispage{\baselineskip}
\clearpage

\begin{figure*}
    \centering
    \includegraphics[width=\textwidth]{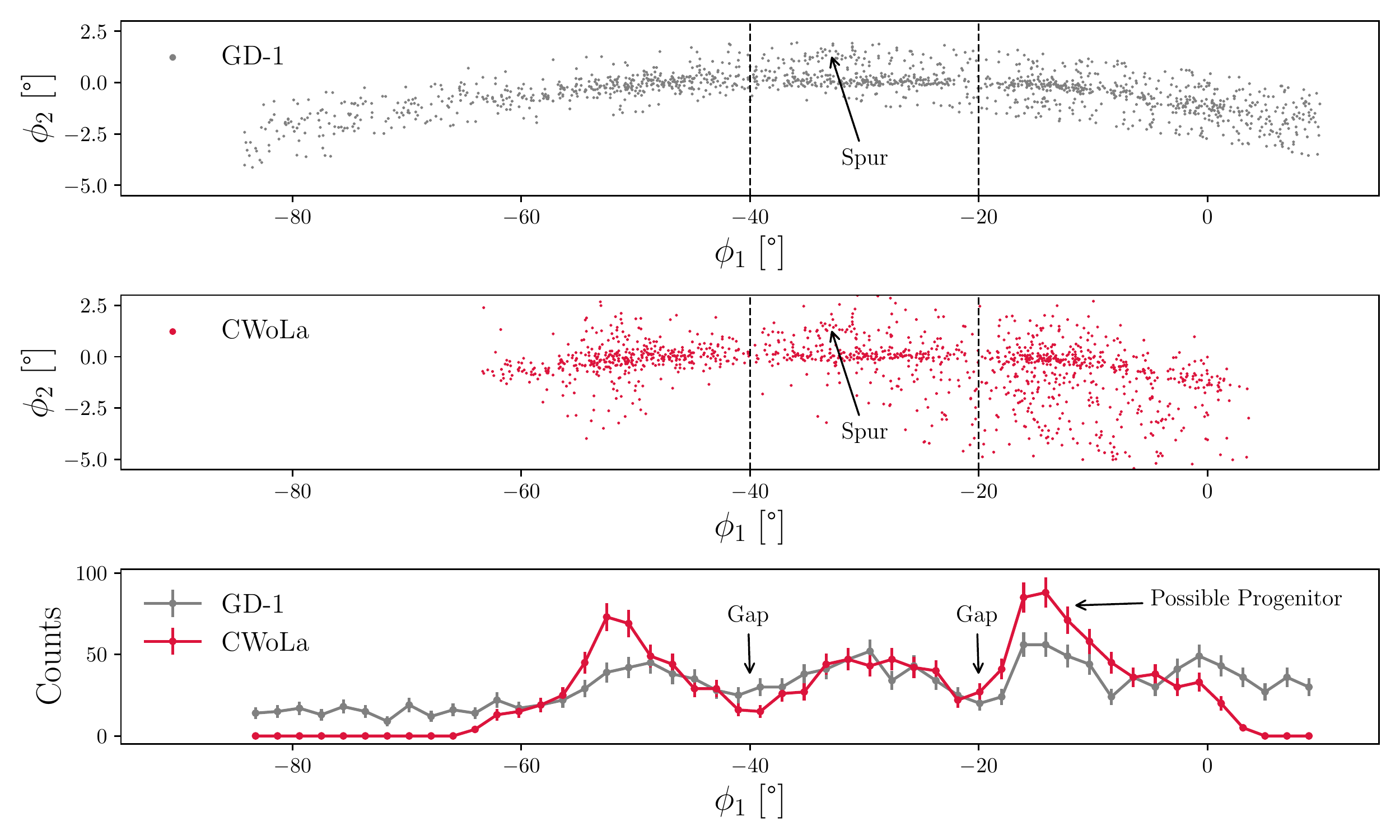}
    \caption{The \textsc{CWoLa}-identified stars across all patches are compared with the labeled GD-1 stars from \cite{pwb18} in stream-aligned coordinates $\phi_1$ and $\phi_2$. This perspective highlights that \textsc{CWoLa} has identified several of the density perturbations unique to GD-1: two sparsely-populated ``gaps'' near $\phi_1 = -40$\textdegree\ and $\phi_1 = -20$\textdegree; an offshoot, or ``spur'', near $\phi_1 = -35$\textdegree; and an overdensity of stars, or ``blob'', near $\phi_1 = -15$\textdegree.} 
    \label{fig:stream_frame}
\end{figure*}

\begin{figure*}
    \centering
    \includegraphics[width=\textwidth]{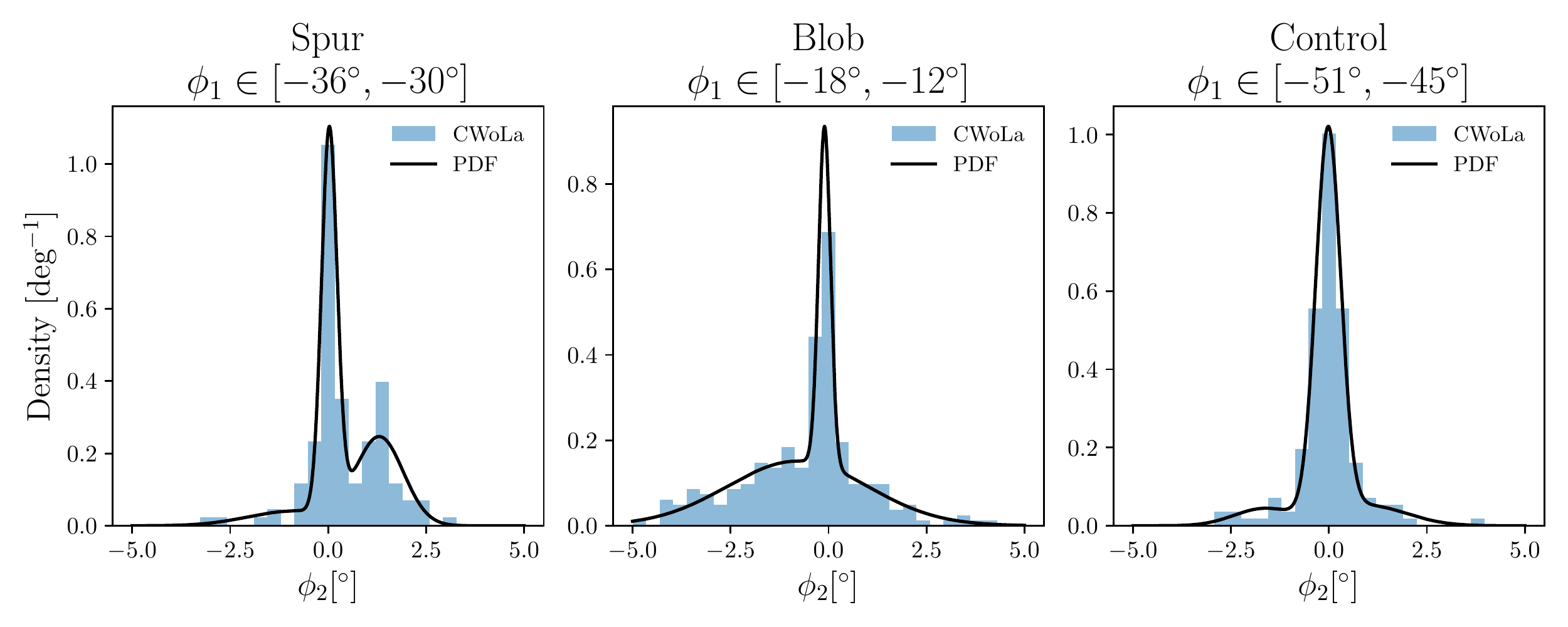}
    \caption{Three subsets of the CWoLa-identified stars (the ``blob'', ``spur'', and a control region) are selected and fit with a three-component Gaussian mixture model to highlight the kinematic qualities of the additional feature, if present. In each case, the GD-1 stream corresponds to the primary narrow peak centered near $\phi_2=0$\textdegree. In the first two plots, we see clear indications of a second peak representing each feature.}
    \label{fig:spurblob}
\end{figure*}

\clearpage

The majority of GD-1 is quite narrow, with an average angular width of approximately 0.5\textdegree \cite{Malhan_2019}, and dense, with approximately 100 stars per 5\textdegree\ bin between $\phi_1 = -60$\textdegree and $\phi_1 = 0$\textdegree. Within this region, \textsc{CWoLa} can reliably identify the stream stars. The tail ends of GD-1 ($\phi_1 \leq -60$\textdegree\ and $\phi_1 \geq 0$\textdegree) are more sparsely populated, with about half the average population per bin of the main body of the stream, and less localized in $\mu_\lambda$, meaning stream stars in this region are harder to identify using \textsc{CWoLa}. Some stars in these regions may also have been excluded from the 21 patches due to their proximity to the Galactic disc or the presence of nearby dust. These regions also tend to include stars with small proper motions, meaning the stream stars are more likely to be overwhelmed by distant background stars.

\begin{figure*}
    \centering
    \begin{subfigure}[t]{0.6\textwidth}
        \centering
    \includegraphics[width=\textwidth]{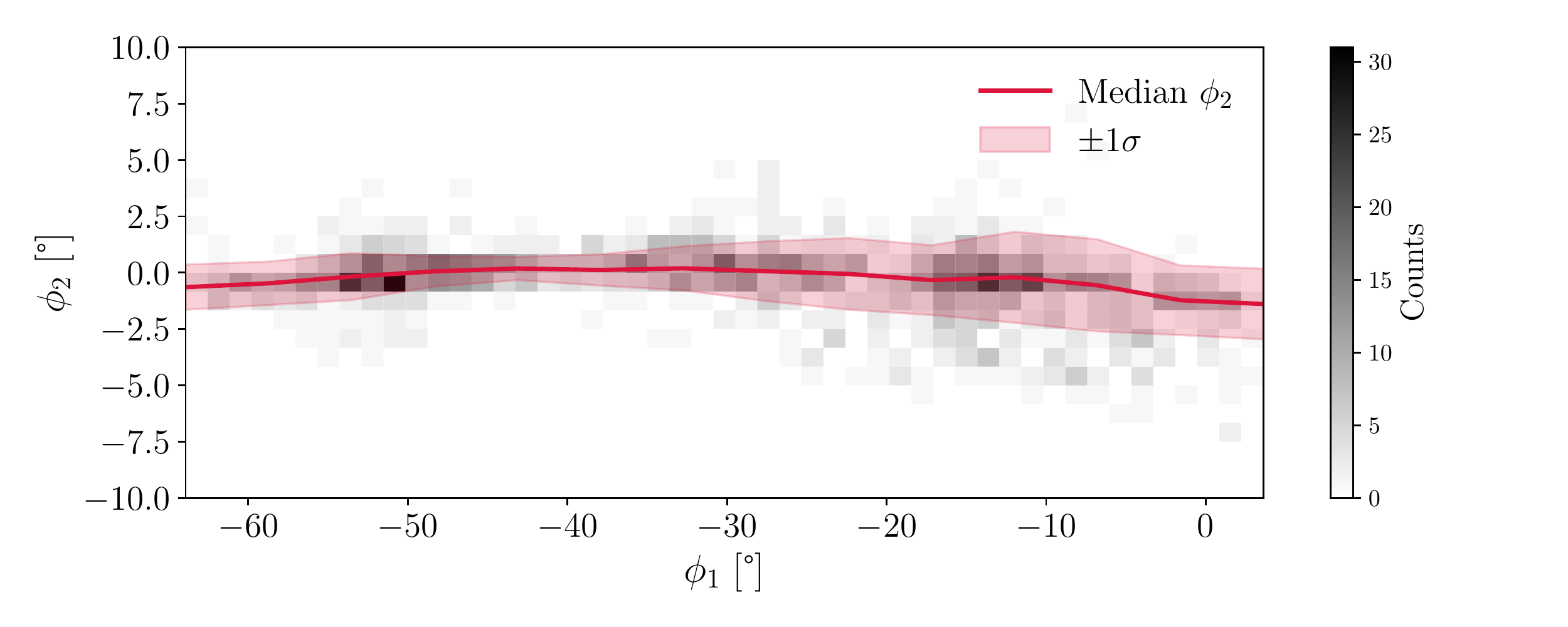}
        \caption{CWoLa-identified stars with the median $\phi_2$ in bins of $\phi_1$}
        \label{fig:median}
    \end{subfigure}\hspace{0.3cm}
    \begin{subfigure}[t]{0.3\textwidth}
        \centering
    \includegraphics[width=0.9\textwidth]{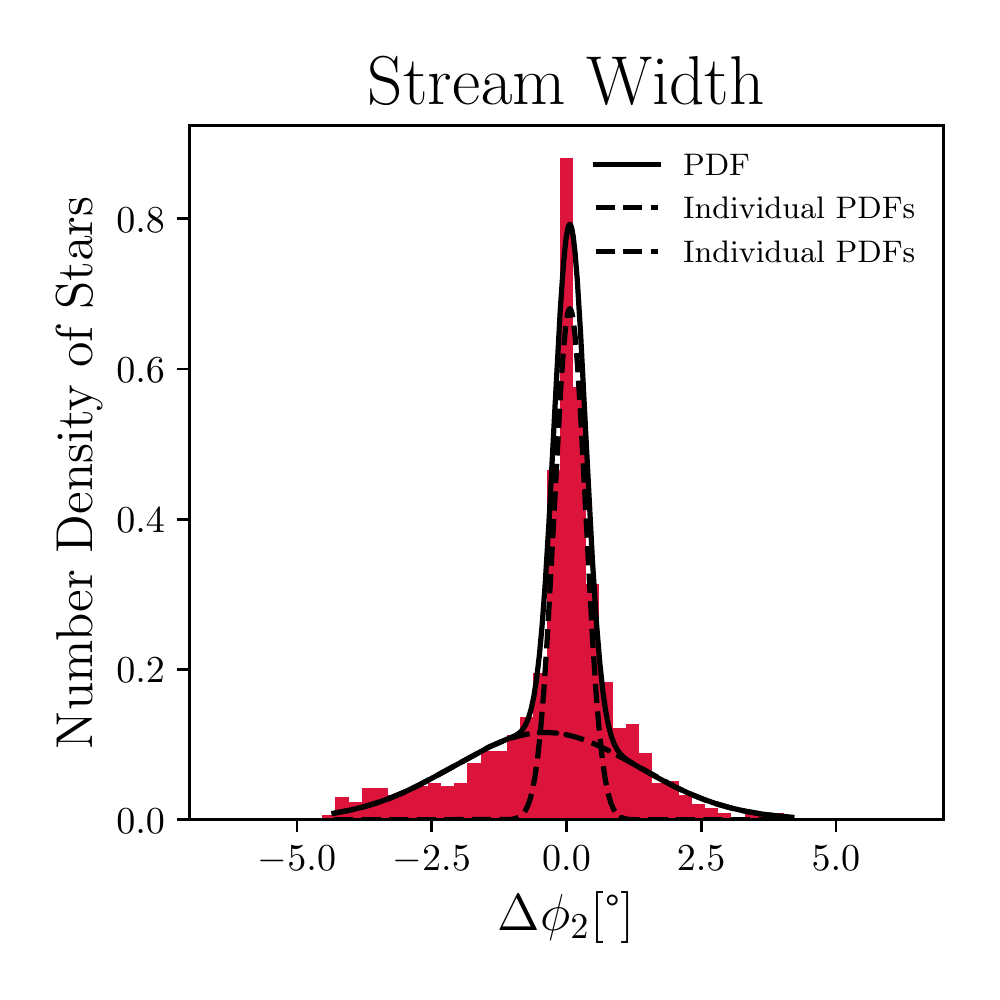}
        \caption{Width of CWoLa stars}
        \label{fig:cocoon}
    \end{subfigure}%
\caption{The width of the CWoLa-identified stars is determined by first calculating the median stream position in $\phi_2$ for 10 bins of $\phi_1$ (the overlaid red line in (a)). The $\phi_2$ coordinates are then shifted by these median values, yielding the histogram in (b). In (b), we use a 2-component Gaussian Mixture Model to show two individual Gaussian components with $\sigma \approx 0.3$\textdegree\ (the core of GD-1) and $\sigma \approx 1.7$\textdegree\ (the ``cocoon''). This appears to support the general trend observed in Figure 6 of \cite{Malhan_2019}.}
\end{figure*}

We can also analyze these results in a rotated set of position coordinates $\phi_1$ and $\phi_2$ \cite{gd1_info} designed to align with the main body of the stream, as shown in Figure \ref{fig:stream_frame}. This perspective highlights that \textsc{CWoLa} has identified several of the density perturbations unique to GD-1: two sparsely-populated ``gaps'' near $\phi_1 = -40$\textdegree\ and $\phi_1 = -20$\textdegree; an offshoot, or ``spur'', near $\phi_1 = -35$\textdegree; and an overdensity of stars, or ``blob'', near $\phi_1 = -15$\textdegree. We can more quantitatively demonstrate the identification of these features, as in Figure 5 of \cite{Price_Whelan_2018}, by looking at histograms of $\phi_2$ in various ranges of $\phi_1$ as shown in Figure \ref{fig:spurblob}. This study highlights the ``spur'' and ``blob'' in particular by fitting a histogram of stars near each feature, along with a third control region, with a three-component Gaussian mixture model assuming a background, the GD-1 stream, and the feature (``blob'' or ``spur''). 

As mentioned above, the underdense regions, or ``gaps'', in GD-1 are typically observed near $\phi_1\approx-40$\textdegree\ and $\phi_1\approx-20$\textdegree. A third underdense region has also been identified near $\phi_1\approx-3$\textdegree\ \cite{deboer}. Our results would not be inconsistent with this third gap, as the density in this area for the \textsc{CWoLa}-identified stars is indeed low, on par with the densities seen at the other two ``gaps'', but since this region is so close to the furthest extent of the \textsc{CWoLa}-identified stars, it is not clear whether this underdensity is a feature from the stream or a reflection of the diminished purity of stars in the patches on the ends of the stream. 

As for the overdense regions, \cite{deboer} reported four overdense regions peaked at $\phi_1\approx-48$\textdegree\ (the highest-density region of the stream), $\phi_1\approx-27$\textdegree, $\phi_1\approx-10$\textdegree, and $\phi_1\approx+2$\textdegree. While the \textsc{CWoLa}-identified stars do not reliably cover the region above $\phi_1=0$\textdegree, the remaining three peaks for which $\phi_1<0$\textdegree\ are also seen in the \textsc{CWoLa}-identified stars. By fitting the \textsc{CWoLa}-identified stars with a mixture model of three Gaussian distributions, we can extract approximate overdensity peaks at $\phi_1\approx-51$\textdegree, $\phi_1\approx-30$\textdegree, and $\phi_1\approx-11$\textdegree. It is interesting to note that \textsc{CWoLa} picks up a large $\phi_1\approx-51$\textdegree\ peak, in line with the highest-density peak reported in \cite{deboer}, though this peak is less pronounced in the stars labeled from \cite{pwb18}.

Another reported feature of GD-1 is a wider ``cocoon'' of stars with a width of around 1\textdegree\ surrounding a much denser core of the stream \cite{Malhan_2019}. To probe this feature, we first calculate the median $\phi_2$ in broad 5\textdegree\ bins of $\phi_1$ to find a smoothed trajectory for the stream. Then, we shift each \textsc{CWoLa}-identified star by its median $\phi_2$ location (see Figure \ref{fig:median}). Once the stream has been centered around this path, we make a 3$\sigma$ selection, as in \cite{Malhan_2019}, then plot the histogram of shifted $\phi_2$ (see Figure \ref{fig:cocoon}). Fitting the distribution to a two-component Gaussian mixture model reveals a narrow peak with a standard deviation of $\sigma\approx 0.3$\textdegree\ (the core of the stream) and an additional wider peak with $\sigma\approx 1.7$\textdegree. This appears to support the observation of such a ``cocoon'' of more diffuse stars surrounding the central core of the stream. 

\begin{figure}[th!]
\centering 
\includegraphics[width=\textwidth]{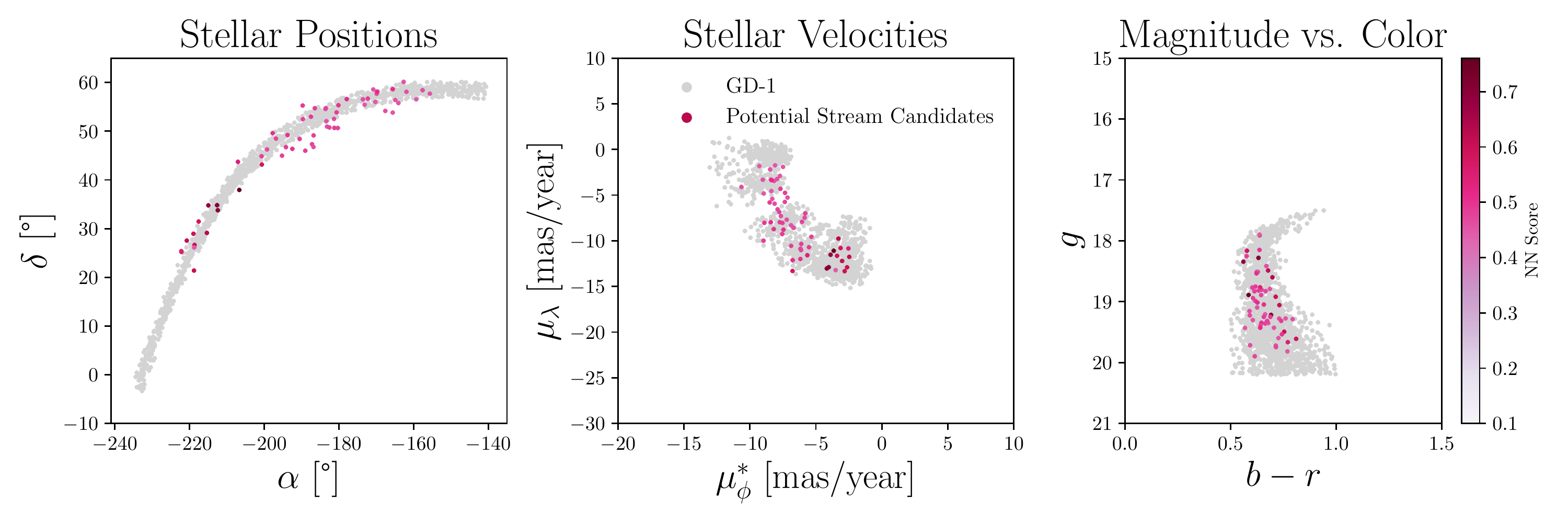}
\caption{\label{fig:promising} Isolating the subset of the top stars identified by \textsc{CWoLa} with the $10\%$ smallest 5-dimensional Euclidean distances $d$ to the nearest labeled star reveals 60 additional stellar candidates for GD-1 membership that may have been omitted from the GD-1 ground truth labeling.}
\end{figure}

\subsection{Towards an Augmentation of the GD-1 Stream Labeling}
\label{ssec:labeling}
Beyond identifying cold stellar streams without labels, the \textsc{CWoLa} technique may also be useful for improving the labeling systems that indicate which stars belong to a particular stream. For instance, \textsc{CWoLa} can identify promising stellar candidates that were not labeled as GD-1 members in \cite{pwb18}, but nevertheless have properties closely aligned with labeled stars. 

As shown in Section \ref{sec:results}, 1,360 unique stars are identified by \textsc{CWoLa} across all 21 patches, and of these, 760 (56\%) are part of the labeled GD-1 star set. We can further refine the remaining 600 unlabeled stars by identifying the subset of individual stars $s$ that minimize the Euclidean distance $d$ to their respective closest labeled GD-1 star in the test set $s'$ along the 5 dimensions used as \textsc{CWoLa} inputs, individually standardized to have $\mu=0$ and $\sigma = 1$: ($\phi$, $\lambda$, $\mu_{\phi *}$, color ($c \equiv b-r$), and magnitude ($g$)): 
\begin{equation}
\label{eq:d}
    d = \sqrt{(\phi - \phi')^2 + (\lambda - \lambda')^2 + (\mu_{\phi *} - \mu'_{\phi *})^2 + (c - c')^2 + (g - g')^2}.
\end{equation}

Isolating the stars yielding the smallest $10\%$ of distances $d$ reveals 60 additional stars, shown in  Figure~\ref{fig:promising} and listed explicitly in Appendix \ref{appendix:promising}, that appear to align with the labeled GD-1 stars across these six dimensions and would be interesting to investigate as potential GD-1 member candidates. A detailed cross-checking of these candidate GD-1 stream stars with other, more precise GD-1 stream catalogs will be pursued in future work. 
\enlargethispage{\baselineskip}
\clearpage

\section{Discussion}
\label{sec:discussion}

It is evident that \textsc{CWoLa} successfully identifies significant portions of GD-1, as measured by not only purity and completeness but also on the faithful reconstruction of physical characteristics and density perturbations characteristic of this stream. Additionally, \textsc{CWoLa} is highly effective at identifying simulated streams. 


While the analysis presented here shares many core strategic components and the same dataset with \textsc{Via Machinae}, this analysis differs primarily in terms of the mechanisms for how to assign anomaly scores to stars, how to divide the sky into subsections for scanning, and how to cluster stars post-training. \textsc{CWoLa} is implemented via a comparatively simple, lightweight, and easy-to-train neural network-based classifier instead of a normalizing flow model to approach the same problem of anomaly detection. When applied to the same example stellar stream, GD-1, \textsc{CWoLa} is able to identify stars with comparable purity with much less computational overhead. We find 760 labeled stars overall, yielding a 56\% purity, while \textsc{Via Machinae}'s first iteration \cite{viamachinae} found 738 stars, yielding a 49\% purity. \textsc{Via Machinae}'s latest iteration \cite{viamachinae2}, which includes additional fiducial selections and an augmented scan over both proper motion variables, increases its star yield to 820, or 65\% purity. 

It is worth emphasizing that our approach does not apply any kind of line-finding or protoclustering algorithms as is done in \textsc{Via Machinae} -- the anomalous stars here are simply combined and filtered via $k$-means clustering. This lightweight clustering strategy is particularly useful in a post-discovery context in which we are interested in refining stream membership catalogues. Another important distinction between these techniques is that \textsc{CWoLa} uses signal and sideband regions of varying widths that are chosen for each patch based on the location of the signal, while \textsc{Via Machinae} searches over regions of interest defined by the orthogonal proper motion coordinate with a fixed width of 6 mas/year with centers spaced 1 mas/year apart. The fiducial selections also differ slightly between these implementations: \textsc{Via Machinae} restricts each patch to the innermost 10\textdegree circle in position space to avoid edge effects, but \textsc{CWoLa} does not exhibit these effects and thus we do not impose this selection. 

Training normalizing flows such as those in \textsc{ANODE} can be a time-intensive task, while completing the full \textsc{CWoLa} training paradigm for GD-1 takes just 15 minutes per patch on an NVIDIA A40 GPU. Running over each patch is embarrassingly parallel and can easily be run simultaneously based on GPU access or using multiprocessing on CPUs. Running all 21 patches on a single GPU takes about 5 hours in total, making it quite feasible for researchers to optimize their signal and sideband region definitions as well as to combine the results from scans of multiple variables. 

This training time can be even further reduced by applying the fiducial selections to the samples \emph{before} training -- cutting training time roughly in half and yielding an overall purity of 44\%. For a 100\% improvement in training time, this technique only reduces the final purity of the identified stars by about 20\%, making this a valuable option particularly for coarse-grained scans across wide areas of phase space. \textsc{CWoLa} and \textsc{Via Machinae} can therefore be thought of as complementary tools for stream detection under different circumstances or computational constraints. 

\begin{figure*}[h!]
    \centering
    \begin{subfigure}[t]{0.45\textwidth}
        \centering
    \includegraphics[width=\textwidth]{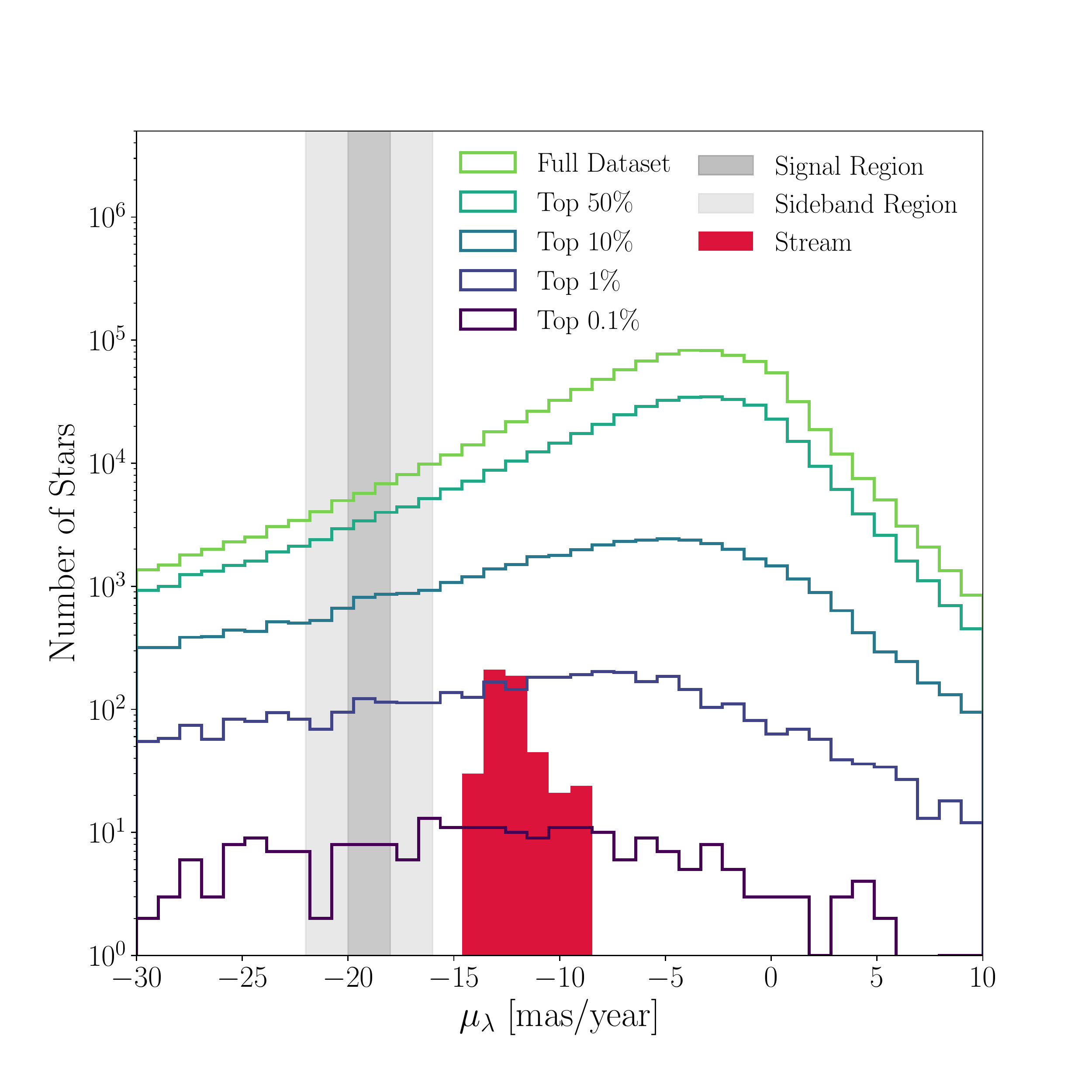}
        \caption{When the signal region does not contain the stream, there is no obvious bump anywhere in proper motion space as the cut percentage gets larger.}
        \label{fig:histo_cut_1}
    \end{subfigure}\hspace{0.5cm}
    \begin{subfigure}[t]{0.45\textwidth}
        \centering
    \includegraphics[width=\textwidth]{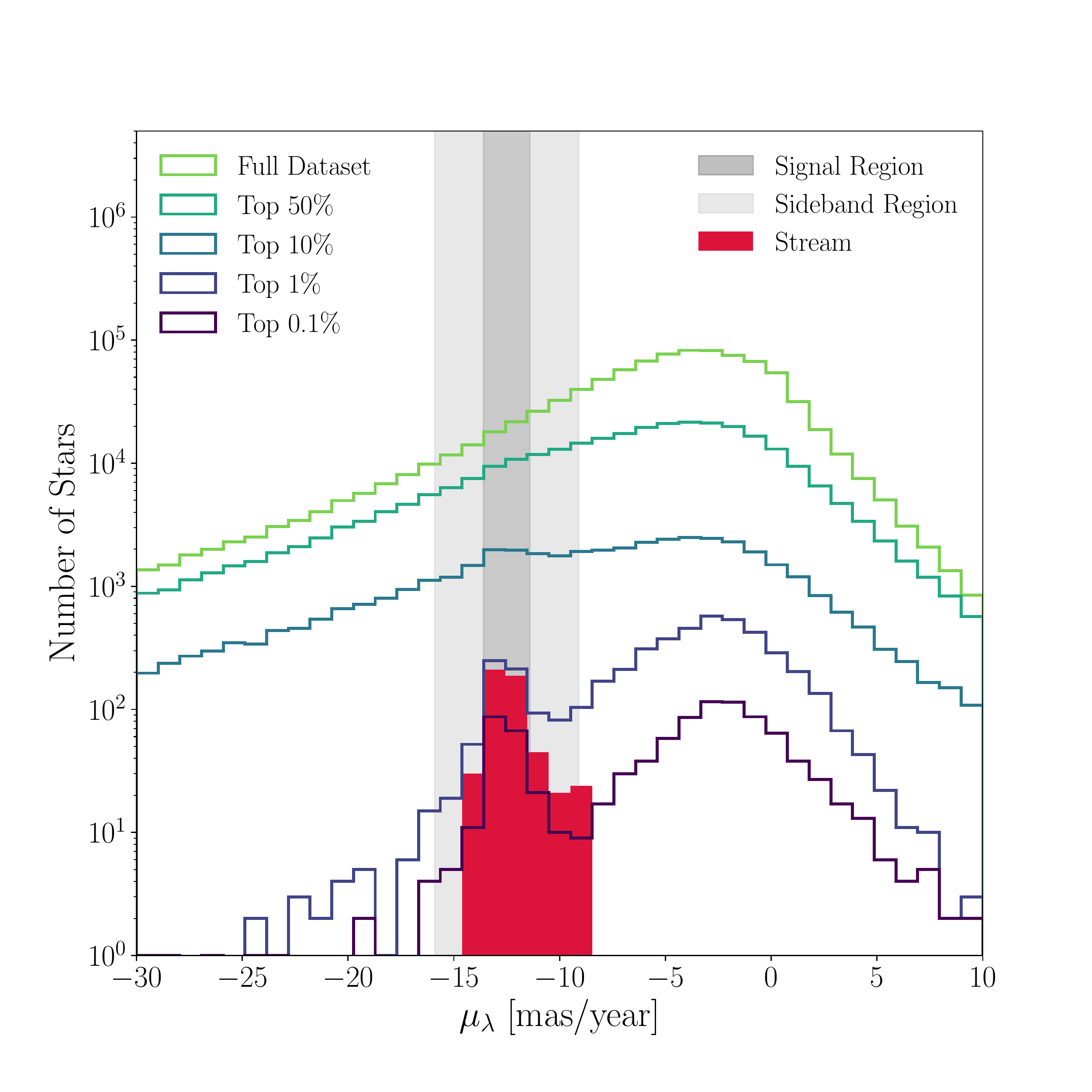}
        \caption{When the signal region contains the stream, an obvious bump forms in the same area where the stream is localized.}
        \label{fig:histo_cut_2}
    \end{subfigure}%
\caption{A demonstration of a scan for which the anomaly location is not previously known.}
\end{figure*}

If \textsc{CWoLa} can be used to identify known streams, it may also be used to potentially find new, undiscovered streams. Some additional challenges will arise when extending \textsc{CWoLa} to look for new stellar streams within the full {\it Gaia} dataset. Detected anomalies are not necessarily guaranteed to be stellar streams, since \textsc{CWoLa} could identify any localized anomalous features. A lack of ground truth labeling for a stream would also require a reevaluation of our performance metrics -- for instance, streams would need to be evaluated using the standard anomaly detection technique of performing a series of selections (e.g. a range of percentiles of the neural network score, or hand-picked thresholds based on the background rate in the sideband region) on a histogram of proper motion. If no anomaly is present, these increasingly harsh selections will reduce the sample statistics without significantly altering the histogram shape, as shown in Figure~\ref{fig:histo_cut_1}. However, if an anomaly is present and identified by \textsc{CWoLa}, a new shape will emerge with increasingly harsh selections on the distribution in question, as shown in Figure~\ref{fig:histo_cut_2}. Additionally, multiple passes of \textsc{CWoLa} might be needed with different choices of signal and sideband region widths if the approximate width of the anomaly is not \emph{a priori} known. 

Searching for new stellar streams will require scanning over the full range of proper motion values in the dataset, since we will not know where new streams might be localized. By applying \textsc{CWoLa} in a coarse sliding-window fashion across $\mu_\lambda$, regions of interest may be identified. These regions can be further studied through finer scans until anomalous data points are identified. When combining multiple patches together for an overall result, we might also need to additionally employ a line-finding algorithm for identifying larger-scale stream-like results, such as the modified Hough transform used in \textsc{Via Machinae} \cite{viamachinae}.

\section{Conclusion}
We have demonstrated a new application of ``\textsc{CWoLa} hunting'', an anomaly detection technique based on the weakly-supervised machine learning classifier \textsc{CWoLa} that is designed to detect localized anomalies in a model-agnostic manner. \textsc{CWoLa} is shown to be easy to train, highly computationally efficient, and, most importantly, effective at identifying anomalies including the stellar stream GD-1 and dozens of simulated streams with high purity. The GD-1 candidate stars identified by \textsc{CWoLa} exhibit the same density perturbations and physical characteristics (the ``spur'', ''blob'', ``cocoon'', gaps, and overdense regions) noted in several independent studies of the stream. The neural network output scores also give clues as to which stars might have been accidentally omitted from more formal GD-1 labeling schemes, suggesting several promising candidates. The successful application of \textsc{CWoLa} in this study shows that \textsc{CWoLa} has strong potential to improve the signal-to-noise ratio on the membership of known streams as well as to potentially reveal previously undetected streams throughout the Galactic halo. \textsc{CWoLa} has broad applicability as a weakly-supervised anomaly detection technique outside of high-energy physics -- and, potentially, it could be applied into still more areas of fundamental science.

\subsection{Reproducing these Results}
A codebase with instructions on how to reproduce each of the plots in this paper is located at \url{https://github.com/hep-lbdl/GaiaCWoLa}. The datasets needed to fully reproduce the plots in this paper (with \textsc{CWoLa} already applied) are publicly available \cite{our_zenodo}. The full 21 patches covering GD-1 are also publicly available \cite{vm_zendodo}. 

\acknowledgments
M.P. thanks Shirley Ho, Adrian Price-Whelan, David Spergel, and Joshua Bloom for their helpful conversations, and gratefully acknowledges the hospitality of the Flatiron Institute Center for Computational Astrophysics. S.T. thanks Vanessa Boehm for her help on working with Optuna for hyperparameter optimization. This work makes use of data from the European Space Agency (ESA) mission
{\it Gaia} (\url{https://www.cosmos.esa.int/gaia}), processed by the {\it Gaia}
Data Processing and Analysis Consortium (DPAC,
\url{https://www.cosmos.esa.int/web/gaia/dpac/consortium}). Funding for the DPAC
has been provided by national institutions, in particular the institutions
participating in the {\it Gaia} Multilateral Agreement. B.N., M. P., and S. T. are supported by the Department of Energy, Office of Science under contract number DE-AC02-05CH11231.

\printbibliography
\appendix
\section{Simulated Stellar Streams}
\label{appendix:mock}
Our implementation of \textsc{CWoLa} for stellar stream discovery was first tested on simulated stellar streams. These simulated streams are frequently highly localized in the proper motion coordinate $\mu_\lambda$, meaning they may tend to be easier to find with our methods than a real stream such as GD-1. 

The streams were simulated using the Gala \cite{gala} Python package to evolve stars in a mock globular cluster along an orbit through the simulated Milky Way potential \cite{milkyway2016potential}, with the center of the stream randomly placed on the sky with a distance randomly chosen between 5 and 20 kpc from the Earth. Stellar properties for the stream components were generated from a MIST \cite{mist1,mist2,mist3,mist4,mist5,mist6} isochrone, assuming [Fe/H] $= -1.6$ and an age of 10 billion years. Observational errors compatible with the Gaia DR2 dataset were added to the synthetic stream stars using the PyGaia \cite{pygaia} package. 

\begin{figure*}[h!]
    \centering
    \includegraphics[width=\textwidth]{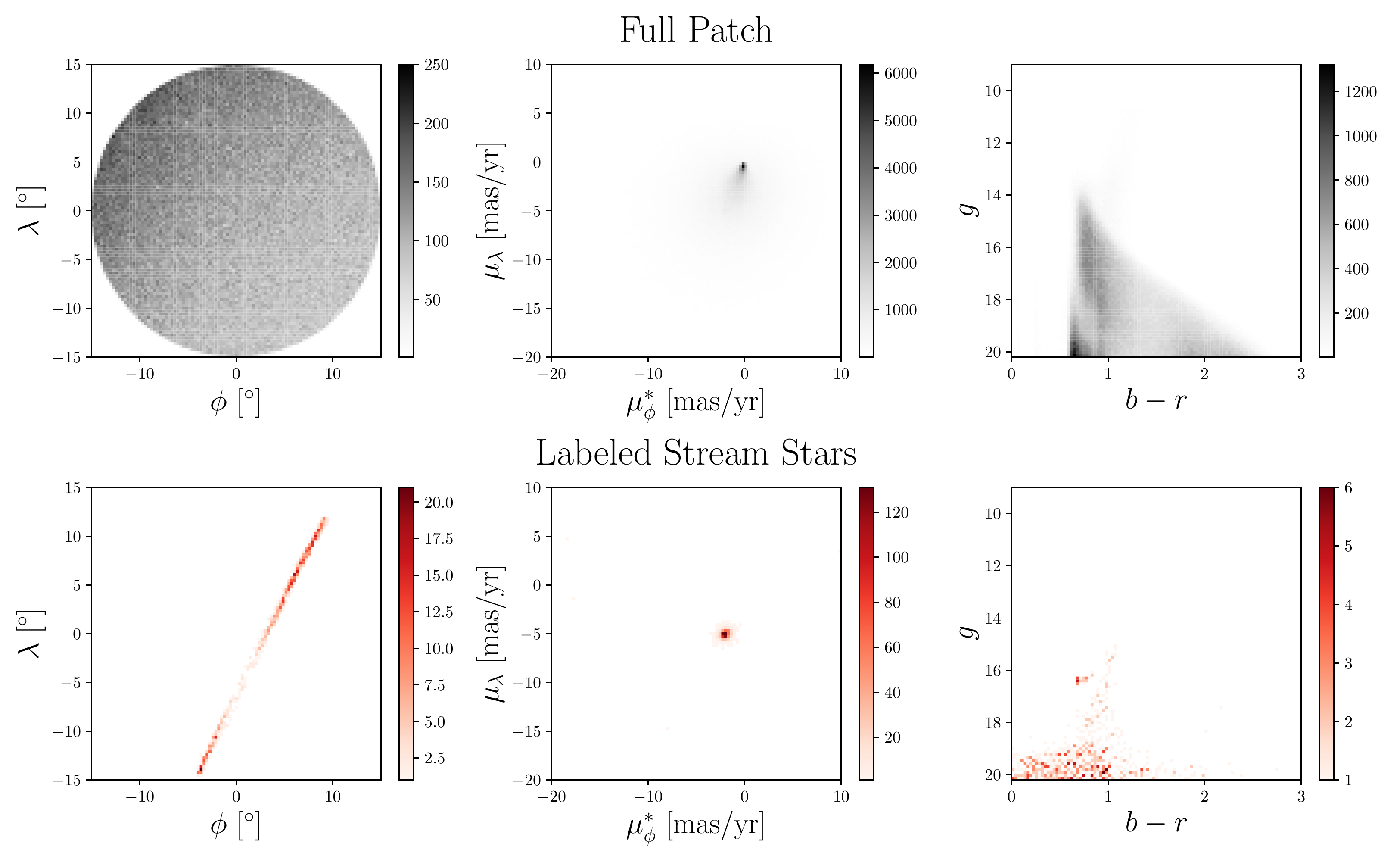}
    \caption{Distributions in position, velocity, and color space for a simulated patch as well as the simulated stream contained within it. While both background stars and simulated stream stars are both highly concentrated in velocity space, the stream stars' peak proper motions are located further from $(\mu_\phi^*, \mu_\lambda) = (0,0)$ than those of the background stars.}
    \label{fig:mock1}
\end{figure*}

An example simulated stream, representing just 1,161 ($0.13\%$) of the 886,677 stars in the simulated patch, is shown in angular position space in Figure~\ref{fig:mock1}. The simulated streams are presented as standalone patches, so \textsc{CWoLA} is applied to just one simulated patch at a time. No fiducial cuts are applied to the simulated patches.

\begin{figure*}[h!]
    \centering
    \includegraphics[width=\textwidth]{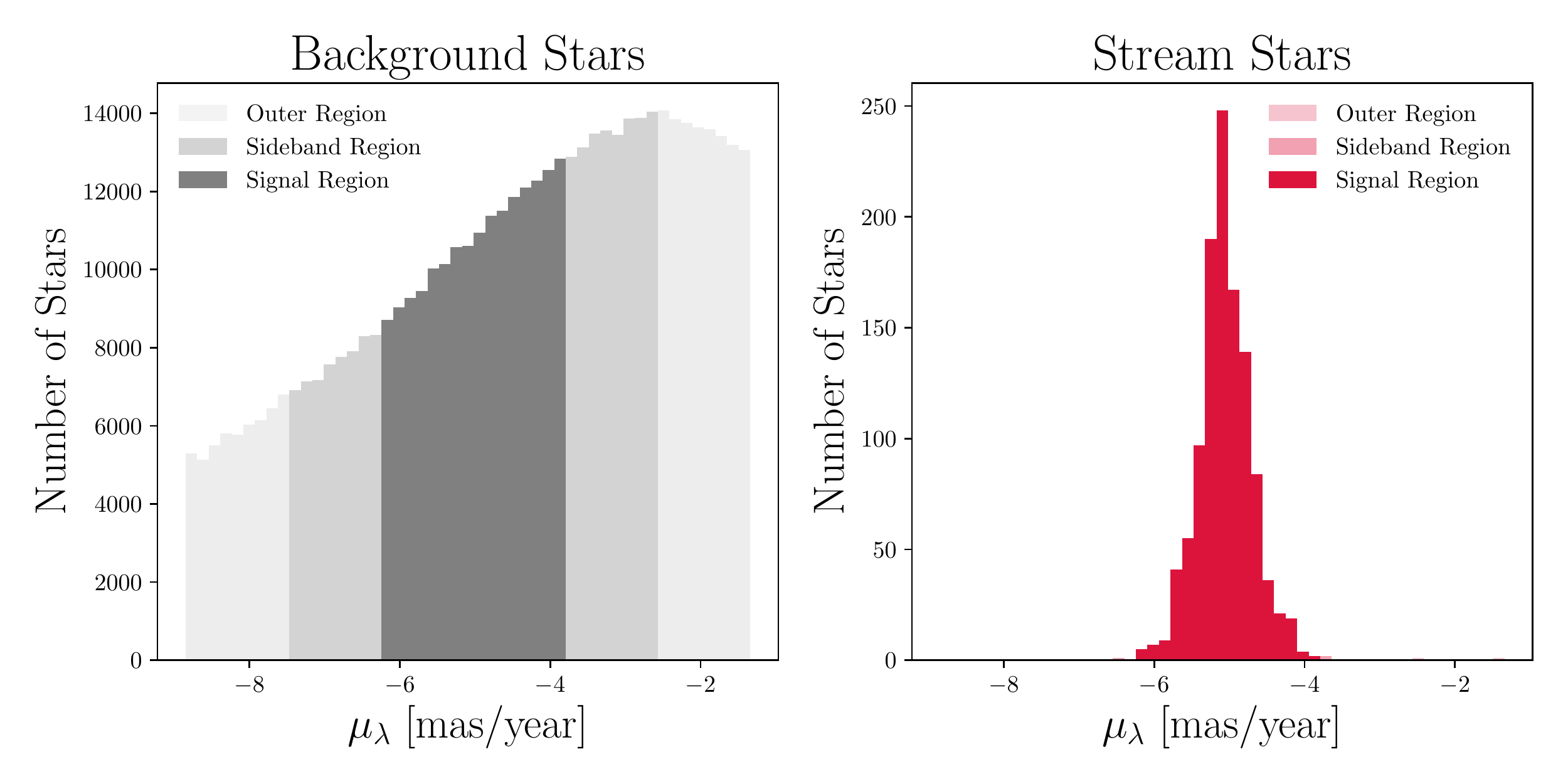}
    \caption{Simulated streams are far more concentrated in angular velocity space than a typical stream in the Gaia dataset. As a result, the signal and sideband regions are defined within a much narrower band around the median stream $\mu_\lambda$. The signal region is defined within $\pm \sigma/4$, or $[-6.3, -3.8]$, while the sideband region is defined as $\pm\sigma/2$, excluding the signal region: $[-7.6, -6.3)\ \&\ (-3.8, 2.6]$. The stream stars are almost exclusively contained within the signal region. }
    \label{fig:mock2}
\end{figure*}

As a proof of concept, we choose idealized signal and sideband limits with prior knowledge of the location of the stellar stream: the sideband is defined as the window of total width $\frac{\sigma}{4}$ surrounding the median $\mu_\lambda$ value of the stream, while the sideband is defined as the additional window of total width $\frac{\sigma}{2}$ surrounding the signal region. These signal and sideband regions for background and stream stars are plotted in Figure \ref{fig:mock2}. 

\begin{figure*}[h!]
    \centering
    \begin{subfigure}[t]{0.45\textwidth}
        \centering
    \includegraphics[width=\textwidth]{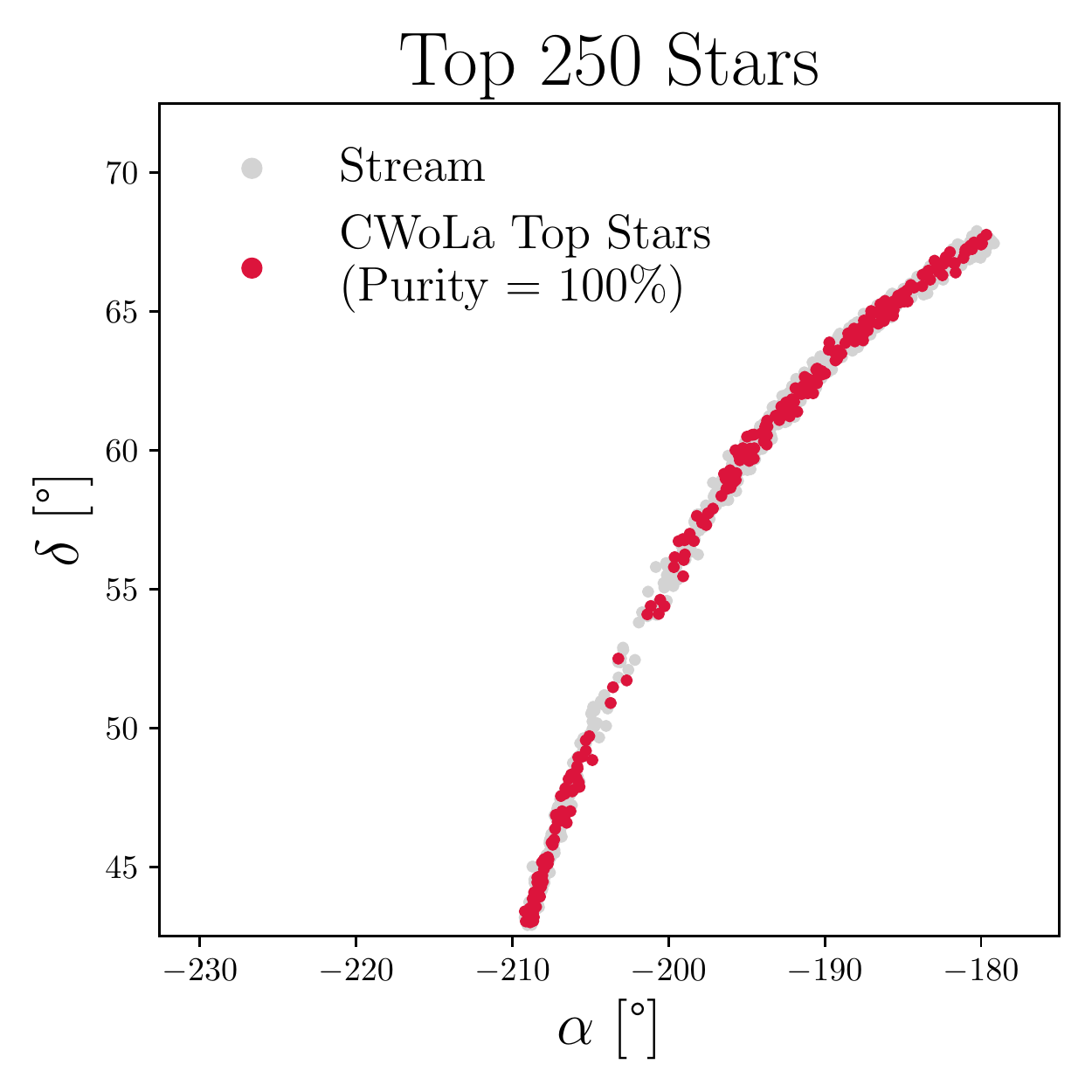}
        \caption{The top 250 stars in one sample simulated stream, as ranked by \textsc{CWoLa} classifier output score, are all members of the ground truth labeled star population.}
    \label{fig:mock_results}
    \end{subfigure}\hspace{0.5cm}
    \begin{subfigure}[t]{0.45\textwidth}
        \centering
    \includegraphics[width=\textwidth]{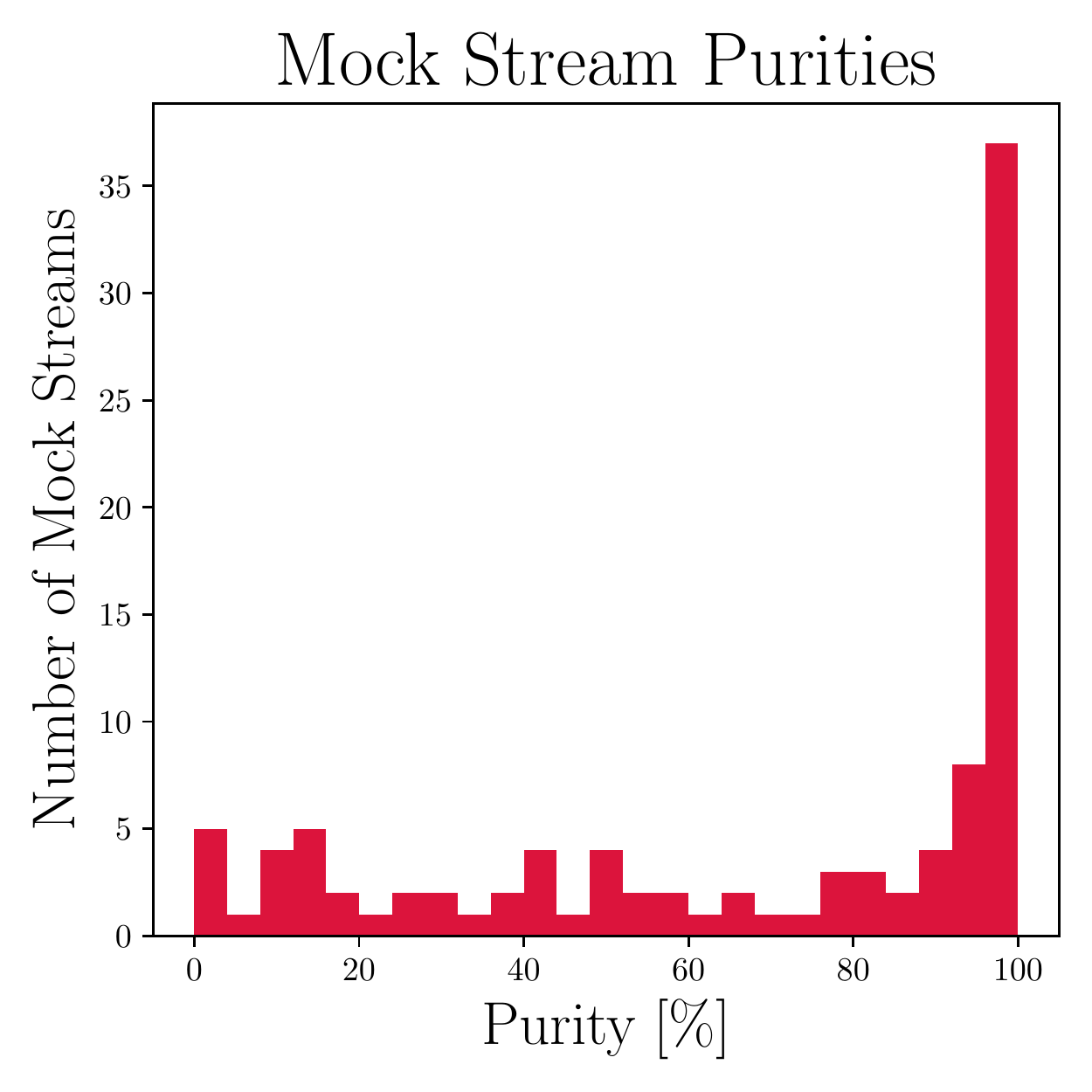}
        \caption{The vast majority of simulated stream stars are identified with high purity levels after two passes of the CWoLa search method.}
        \label{fig:mock_purities}
    \end{subfigure}%
\caption{CWoLa performance evaluated as a function of purity across multiple simulated streams.}
\label{fig:histo_cut}
\end{figure*}

We train \textsc{CWoLa} to distinguish between events from these signal and sideband regions, then select the top 250 stars as ranked by CWoLa's classifier output score. As shown in Figure~\ref{fig:mock_results}, 100\% of the top 250 stars selected for this patch are members of the ground truth labeled stream population in this patch. 

When this technique is applied across 100 randomly-sampled simulated streams, 76\% of streams are identified with purity $> 0$\%, of which 75\% are identified with high purity (defined as purity greater than 50\%). However, a large portion of these cases with zero purity are streams with wider distributions along $\mu_\lambda$, so the results can be further augmented with additional scans choosing different signal and sideband regions. When supplemented with an additional scan with wider signal and sideband region definitions (signal region = $\pm \sigma$ and sideband region = $\pm 3\sigma$ $-$ signal region), 96\% of streams are identified with nonzero purity, of which 69\% are identified with high purity. Across the 100 streams, the median purity of the \textsc{CWoLa}-identified results is 86\%. Figure \ref{fig:mock_purities} illustrates that the clear majority of the simulated streams are identified with high purity levels.

\clearpage
\section{Patch-by-Patch Performance}
\label{appendix:patches}

\begin{figure*}[h]
    \centering
    \includegraphics[width=\textwidth]{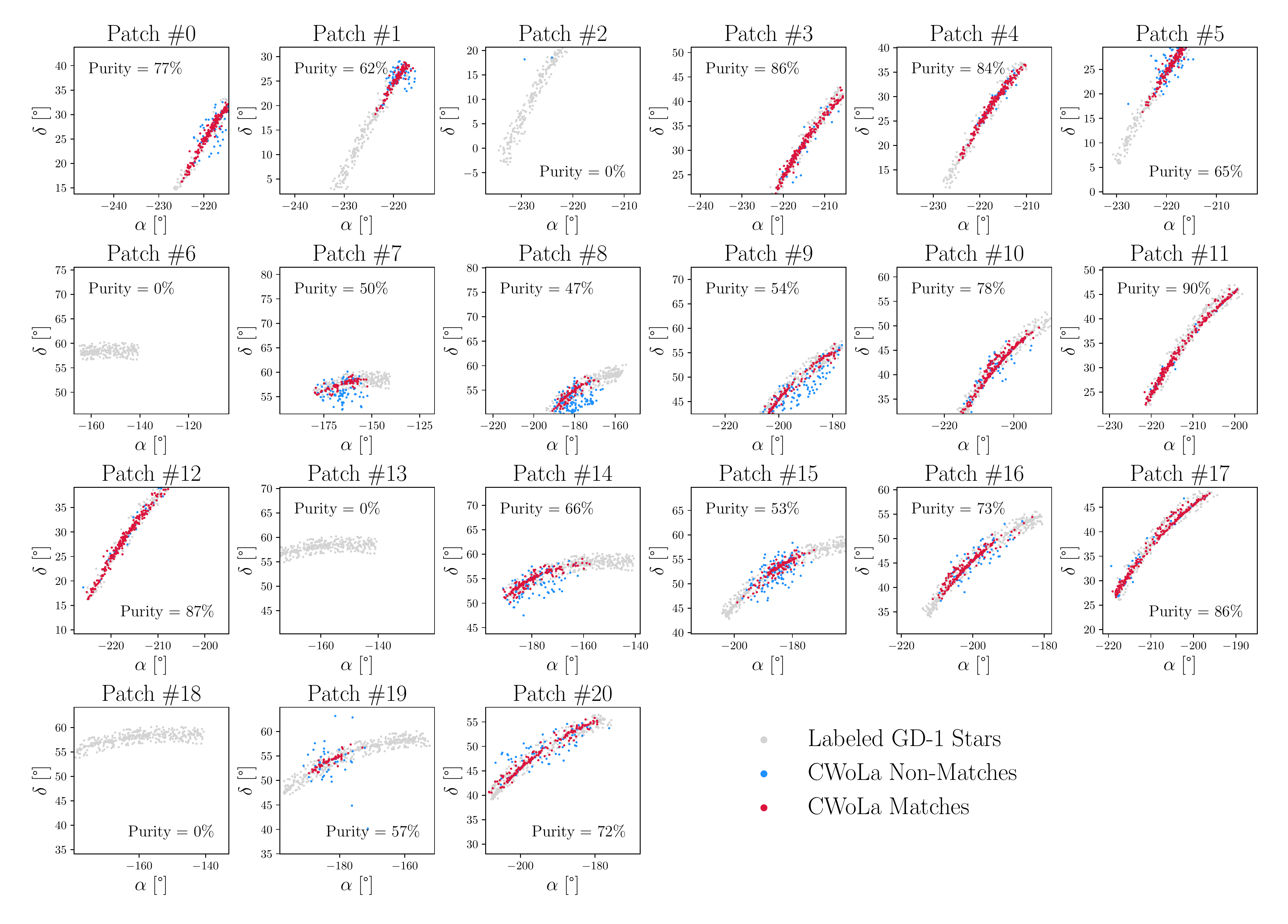}
    \caption{The top 250 identified stars across each of the patches of GD-1 from the Gaia dataset show that CWoLa is able to effectively identify GD-1 stars with high purity levels across all patches, with the exception of patches on the very furthest tails of the stream.}
    \label{fig:all_patches}
\end{figure*}

Figure \ref{fig:all_patches} shows the patch-by-patch breakdown of \textsc{CWoLa} applied to GD-1. Each of the 21 patches is considered separately for individual applications of the \textsc{CWoLa} methodology, including fiducial cuts and $k$-means clustering. These results are combined in Figure \ref{fig:rainbow}. \textsc{CWoLa} achieves a high purity across nearly all patches, with the exception of those patches with relatively fewer stream stars located at the leftmost and rightmost edges of the stream (near $\alpha = -230$\textdegree and $\alpha = -150$\textdegree).

\clearpage
\section{Potential GD-1 Candidate Members}
\label{appendix:promising}

The following table details the subset of unlabeled stars identified by CWoLa that fall within the smallest 10\% of five-dimensional distances $d$ (see Equation \ref{eq:d}) to stars in the labeled GD-1 set, ranked in descending order by CWoLa's neural network (NN) classifier score. 

\bigskip

\centering
\scalebox{0.75}{
\small
\begin{tabular}{lrrrrrrrrr}
Index &  Patch &   $\alpha$ &     $\delta$ &    $\mu_{\phi}^*$ &        $\mu_\lambda$ &       $b-r$ &          $g$ &  $d$ &  NN Score \\
0  &         3 &  153.343750 &  37.944386 &  -3.648079 & -11.087369 &  0.585669 &  18.889666 &     0.199219 &  0.760671 \\
1  &         3 &  147.407532 &  34.807056 &  -4.174263 & -13.030541 &  0.632448 &  18.280497 &     0.161793 &  0.710143 \\
2  &         3 &  147.581451 &  33.771442 &  -4.031364 & -12.893708 &  0.692181 &  19.217056 &     0.200405 &  0.703576 \\
3  &         3 &  145.067444 &  34.774712 &  -3.889829 & -11.528839 &  0.560968 &  18.343340 &     0.208592 &  0.687666 \\
4  &         0 &  144.668640 &  29.123871 &  -3.294825 &  -9.761044 &  0.698603 &  18.599613 &     0.157535 &  0.628063 \\
5  &         0 &  139.284058 &  27.534752 &  -2.474001 & -11.759004 &  0.678160 &  18.488266 &     0.202521 &  0.608509 \\
6  &         0 &  141.217377 &  21.388325 &  -3.016653 & -12.211497 &  0.731615 &  19.056505 &     0.207587 &  0.607377 \\
7  &         0 &  141.366760 &  26.641132 &  -2.638692 & -12.904817 &  0.754404 &  19.493120 &     0.169499 &  0.603854 \\
8  &         0 &  137.850464 &  25.221642 &  -2.823553 & -13.327346 &  0.713997 &  18.919308 &     0.181071 &  0.600161 \\
9  &         0 &  141.072662 &  28.944424 &  -3.398272 & -11.644784 &  0.810871 &  19.610243 &     0.077491 &  0.590473 \\
10 &        10 &  159.375610 &  43.155155 &  -6.771805 & -13.304443 &  0.577917 &  18.163719 &     0.193080 &  0.582036 \\
11 &         0 &  142.465027 &  31.453848 &  -3.129723 & -10.795928 &  0.638067 &  17.908943 &     0.162005 &  0.582026 \\
12 &         0 &  137.821594 &  25.365973 &  -2.537415 & -10.836835 &  0.771456 &  19.665091 &     0.186863 &  0.547747 \\
13 &        10 &  152.977966 &  43.714725 &  -5.651521 & -11.580858 &  0.640087 &  18.764666 &     0.173924 &  0.541810 \\
14 &        14 &  190.253159 &  58.102585 &  -8.347640 &  -3.359662 &  0.640982 &  19.427656 &     0.206074 &  0.513133 \\
15 &        14 &  194.429871 &  58.634842 &  -8.420991 &  -7.967748 &  0.623426 &  18.534977 &     0.183633 &  0.506830 \\
16 &        15 &  182.109512 &  56.585564 &  -6.765204 & -12.116959 &  0.627577 &  19.014071 &     0.191147 &  0.505615 \\
17 &        16 &  162.275238 &  49.602287 &  -7.460149 &  -8.779883 &  0.658072 &  19.046940 &     0.198527 &  0.505006 \\
18 &        16 &  166.262054 &  49.207397 &  -8.196286 &  -8.721247 &  0.743465 &  19.535479 &     0.206786 &  0.498103 \\
19 &         9 &  167.548462 &  46.377983 &  -6.176190 & -11.992228 &  0.646498 &  19.345844 &     0.088779 &  0.497648 \\
20 &        14 &  179.886322 &  55.335400 &  -8.907069 &  -8.033201 &  0.637199 &  18.149529 &     0.158184 &  0.492758 \\
21 &        15 &  172.536072 &  52.980537 &  -7.740779 &  -7.965540 &  0.740067 &  19.315273 &     0.179650 &  0.487783 \\
22 &        15 &  170.305725 &  55.276653 &  -8.972501 &  -9.987629 &  0.618397 &  18.751921 &     0.121189 &  0.484890 \\
23 &         8 &  178.793396 &  50.679008 &  -7.688651 &  -4.308014 &  0.688902 &  19.250017 &     0.209937 &  0.484879 \\
24 &        19 &  179.387024 &  53.861988 &  -7.553021 &  -9.258564 &  0.633465 &  18.817064 &     0.177797 &  0.484814 \\
25 &        14 &  173.568787 &  54.718918 &  -8.506504 &  -5.805184 &  0.657850 &  19.249598 &     0.127450 &  0.484371 \\
26 &         9 &  169.494263 &  48.418388 &  -6.111476 & -10.342608 &  0.673502 &  19.317274 &     0.128495 &  0.484024 \\
27 &         8 &  187.757996 &  56.693752 &  -7.339619 &  -4.749725 &  0.576080 &  18.252359 &     0.151545 &  0.484008 \\
28 &         9 &  165.833618 &  46.725655 &  -5.527166 & -10.702893 &  0.613531 &  18.830116 &     0.150776 &  0.483593 \\
29 &         9 &  171.009125 &  45.997330 &  -5.812016 &  -6.994314 &  0.617441 &  18.986839 &     0.169053 &  0.483200 \\
30 &         9 &  173.235535 &  49.141151 &  -5.340712 &  -9.560485 &  0.606213 &  18.940699 &     0.100418 &  0.479875 \\
31 &         8 &  179.718201 &  50.653923 &  -7.778558 &  -6.846837 &  0.664310 &  19.358950 &     0.182553 &  0.477425 \\
32 &         9 &  159.310638 &  44.860550 &  -6.879270 & -10.555117 &  0.646944 &  18.803308 &     0.209683 &  0.477115 \\
33 &         9 &  164.777161 &  44.976212 &  -6.129719 & -11.014638 &  0.603947 &  18.769825 &     0.195550 &  0.476689 \\
34 &         9 &  176.500488 &  54.726501 &  -7.533120 &  -8.043803 &  0.614882 &  19.896984 &     0.196763 &  0.475622 \\
35 &        19 &  177.458374 &  50.769260 &  -7.356768 &  -5.745450 &  0.645157 &  18.890451 &     0.145692 &  0.475277 \\
36 &         9 &  172.817383 &  47.330818 &  -6.053680 &  -7.922386 &  0.642633 &  19.386862 &     0.202310 &  0.474907 \\
37 &         9 &  163.159119 &  48.486298 &  -6.168635 & -10.865490 &  0.666342 &  18.826414 &     0.182255 &  0.472844 \\
38 &        15 &  170.374420 &  52.475494 &  -7.904741 &  -6.552203 &  0.715446 &  19.745758 &     0.196269 &  0.472336 \\
39 &         9 &  160.771454 &  46.261520 &  -5.855455 &  -7.474462 &  0.591482 &  19.225233 &     0.171725 &  0.464586 \\
40 &         7 &  186.439041 &  56.541378 & -10.644468 &  -4.102360 &  0.592939 &  19.715214 &     0.175948 &  0.463402 \\
41 &         8 &  178.696075 &  52.531986 &  -7.652631 &  -7.286445 &  0.669975 &  18.414928 &     0.178466 &  0.462668 \\
42 &         7 &  190.129303 &  57.698757 &  -8.130330 &  -5.936471 &  0.590197 &  19.154623 &     0.140599 &  0.462299 \\
43 &         8 &  176.834106 &  50.947758 &  -7.213428 &  -7.684932 &  0.678570 &  19.351543 &     0.182765 &  0.461110 \\
44 &         9 &  173.211731 &  46.760998 &  -6.690597 &  -8.577130 &  0.730625 &  19.276901 &     0.171330 &  0.460547 \\
45 &         7 &  202.413101 &  58.419315 &  -8.423603 &  -3.303266 &  0.626270 &  19.095215 &     0.075321 &  0.460429 \\
46 &         7 &  198.092728 &  58.096100 &  -7.937617 &  -3.228269 &  0.629757 &  18.509893 &     0.170395 &  0.459476 \\
47 &         8 &  176.415344 &  54.566105 &  -7.152437 &  -5.265400 &  0.794621 &  19.286375 &     0.115260 &  0.458131 \\
48 &         7 &  192.434601 &  54.163914 &  -9.025725 &  -3.314333 &  0.714384 &  19.720222 &     0.199648 &  0.457166 \\
49 &         7 &  197.343521 &  60.127289 &  -8.470557 &  -2.185798 &  0.705921 &  19.428307 &     0.190847 &  0.456135 \\
50 &         8 &  176.661682 &  52.064625 &  -6.882434 &  -8.300988 &  0.686596 &  18.789845 &     0.134871 &  0.454971 \\
51 &         7 &  194.420959 &  53.819744 &  -7.492790 &  -1.907870 &  0.605728 &  19.301640 &     0.188744 &  0.452361 \\
52 &         7 &  195.111771 &  56.396278 &  -8.144449 &  -3.451988 &  0.727108 &  19.597523 &     0.174442 &  0.452323 \\
53 &         7 &  204.361725 &  57.704208 &  -8.309439 &  -5.412242 &  0.568638 &  19.430788 &     0.205758 &  0.452242 \\
54 &         7 &  195.939850 &  55.778358 &  -7.729141 &  -2.890564 &  0.624128 &  18.510710 &     0.162695 &  0.451727 \\
55 &         7 &  189.208954 &  58.567673 &  -8.094695 &  -1.741354 &  0.663921 &  19.207617 &     0.192408 &  0.449935 \\
56 &         7 &  186.898468 &  55.441170 &  -8.953770 &  -4.824331 &  0.760971 &  19.275148 &     0.133778 &  0.448510 \\
57 &         7 &  189.780121 &  56.009251 &  -9.288628 &  -1.816544 &  0.769598 &  19.816544 &     0.209066 &  0.447054 \\
58 &         7 &  200.713043 &  56.548923 &  -8.371297 &  -4.555718 &  0.603445 &  18.767347 &     0.129104 &  0.446953 \\
59 &         1 &  141.269653 &  26.119368 &  -3.506178 & -13.209450 &  0.637341 &  17.898008 &     0.194991 &  0.444487 \\
\end{tabular}
}










\end{document}